\documentclass[11pt,letterpaper]{article}
\usepackage{soul}
\usepackage{jheppub}
\usepackage{ucs}
\usepackage[utf8]{inputenc}
\usepackage{CJKutf8}
\usepackage[T1]{fontenc}
\usepackage{amsfonts}
\usepackage{amsmath}
\usepackage{graphicx}
\usepackage{subcaption}
\usepackage{bm}
\usepackage{tensor} 
\usepackage{mathtools} %
\allowdisplaybreaks[4]
\usepackage{amssymb}
\usepackage{amsthm,amsmath,amssymb}
\usepackage{mathrsfs}
\usepackage{euscript}
\usepackage{color}

\newcommand{\deleted}[1]{\textcolor{red}{#1}}
\usepackage{braket}
\usepackage{orcidlink}
\usepackage{placeins} 
\usepackage{indentfirst}
\usepackage{adjustbox}

\title{Charged Black Holes with a Lorentz--Violating Kalb--Ramond Background}

\author[a]{Jia-Hui Yang\orcidlink{0000-0002-8179-9365}$^{\dagger}$,}
\emailAdd{yjiahui2024@lzu.edu.cn}

\author[a]{Xin-Yu Guo\orcidlink{0000-0001-6670-7955}$^{\dagger}$,}
\emailAdd{xyg2023@lzu.edu.cn}

\author[a,b,c]{Jia-Zhou Liu\orcidlink{0000-0001-6670-7955},}
\emailAdd{liujzh2025@lzu.edu.cn}

\author[a,b,c]{Yu-Xiao Liu\orcidlink{0000-0002-4117-4176}\footnote{Corresponding author}}
\emailAdd{liuyx@lzu.edu.cn}

\affiliation[a]{School of Physical Science and Technology, Lanzhou University, Lanzhou 730000, China \vspace{0.1cm}}

\affiliation[b]{Lanzhou Center for Theoretical Physics, Key Laboratory of Theoretical Physics of Gansu Province, Key Laboratory of Quantum Theory and Applications of MoE, Gansu Provincial Research Center for Basic Disciplines of Quantum Physics, Lanzhou University, Lanzhou 730000, China\vspace{0.1cm}}

\affiliation[c]{ Institute of Theoretical Physics \& Research Center of Gravitation, Lanzhou 
University, Lanzhou 730000, China \vspace{0.1cm}}

\abstract{We investigate exact static, spherically symmetric electrically charged black hole solutions in a gravitational theory with spontaneous Lorentz-symmetry breaking induced by a background Kalb--Ramond field. In contrast to previous analyses that retained only one nonminimal curvature coupling, we include the combined effects of the two independent nonminimal curvature couplings and obtain charged black hole solutions both with and without a cosmological constant. Using the Iyer--Wald covariant phase-space formalism, we derive the corrected thermodynamic quantities and analyze the Joule--Thomson expansion, including the inversion curve and the cooling/heating regions. We further apply the topological approach to black hole thermodynamics to characterize the van der Waals-like phase transition and show how the thermodynamic critical temperature and pressure are encoded in the corresponding topological defect curve. These results clarify the thermodynamic and topological signatures of electrically charged black holes in gravity with a Lorentz-violating Kalb--Ramond background.
}

\begin{document}
\maketitle

\flushbottom

\section{Introduction}\label{sec:intro}
A consistent unification of general relativity and quantum theory remains one of the central goals of modern theoretical physics. Many candidate theories of quantum gravity are formulated near the Planck scale, $10^{19}\,{\rm GeV}$, where direct experimental access is currently unavailable. Nevertheless, several approaches, including string theory \cite{a1}, loop quantum gravity \cite{a2}, Ho\v{r}ava--Lifshitz gravity \cite{a3}, and noncommutative field theory \cite{a4}, indicate that Lorentz symmetry may be broken in the low-energy effective description of the gravitational sector. Such effects provide a phenomenological window into possible microscopic structures of spacetime.

The Standard-Model Extension (SME) provides a systematic effective-field-theory framework for parametrizing Lorentz-symmetry violation and its possible couplings to matter and gravity \cite{a5}.In this context, the bumblebee model utilizes the nonminimal coupling of the bumblebee field $B_{\mu}$ to gravity to generate spontaneous Lorentz symmetry breaking effects\cite{by1,by3,by4,by5,by7,by9,by24,by25,by26,by29,by30}. The Kalb–Ramond antisymmetric tensor arises naturally in the bosonic spectrum of string theory, and a nonzero vacuum expectation value of $B_{\mu\nu}$ selects preferred spacetime directions. In the gravitational sector considered here, spontaneous Lorentz-symmetry breaking is implemented through the vacuum configuration of a Kalb--Ramond two-form field $B_{\mu\nu}$, whose effects are transmitted to the metric by nonminimal curvature couplings.  Kalb--Ramond backgrounds have been used to construct black hole solutions \cite{a6,a7,DZY23,by6,by28}, traversable wormholes \cite{a8,a9}, and to study black hole shadows and quasinormal modes \cite{a10,by2,by10,by23,by27}. Most existing analyses, however, retained only the coupling $B^{\rho\mu}B^\nu{}_\mu R_{\rho\nu}$. Following Ref. \cite{LWWL25}, we keep both independent nonminimal couplings, namely $\xi_1 B^{\mu\nu}B_{\mu\nu}R$ and $\xi_2 B^{\rho\mu}B^\nu{}_\mu R_{\rho\nu}$, and examine their combined effect in the electrically charged sector.

Black hole thermodynamics offers a powerful framework for probing gravitational dynamics beyond the purely geometric description. Since the seminal work of Bekenstein and Hawking \cite{t11,t12,t14,t15,t16}, it has revealed deep connections among general relativity, thermodynamics, and quantum theory. Black holes also display rich phase structures, including transitions analogous to those of ordinary thermodynamic systems \cite{a12,a13,a14,a15,a16,a17,a18,a19,a20,a21,a22,a23,a24,a25,a26,a27,a28,a29,a30}. More recently, Refs. \cite{Wei:2022dzw,Wei:2024gfz,Wei2026} developed a topological description of black hole phase transitions. In this approach, black hole branches are identified with defects of a thermodynamic vector field in Duan's topological current theory and are assigned winding numbers. The total topological number then provides a robust way to classify phase-transition patterns \cite{Wei:2024gfz,Wei2026,by11,by12,by13,by14,by15,by16,by17}.

Electric charge substantially modifies both the causal structure and the thermodynamic behavior of black holes. In anti-de Sitter(AdS) spacetime, charged black holes exhibit thermodynamic features closely analogous to those of van der Waals fluids \cite{OA17,by21,by22}. The purpose of this work is therefore twofold: first, to construct electrically charged, static, spherically symmetric black hole solutions in the presence of the two Kalb--Ramond curvature couplings; and second, to analyze their thermodynamic and topological properties. In the extended phase space, we obtain the Joule--Thomson inversion curve separating cooling and heating regions in the $T$-$P$ plane and recover the Reissner--Nordstr\"om--AdS ratio $T_i^{\min}/T_c=1/2$ \cite{Mo:2018ddimchargedjt,Media:2025kerrdslorentzviolationjt}. We also use the topological method to show how the defect curve encodes the critical temperature and pressure associated with the van der Waals-like transition.

The paper is organized as follows. In Sec. \ref{sec:field-equations}, we introduce the Kalb--Ramond field as a source of spontaneous Lorentz-symmetry breaking and derive the gravitational field equations. In Sec. \ref{sec:solutions}, we present electrically charged, static, spherically symmetric black hole solutions with and without a cosmological constant. Section \ref{sec:thermodynamics} is devoted to the Iyer--Wald thermodynamic analysis and to the Joule--Thomson expansion. In Sec. \ref{sec:topology}, we review the topological framework for black hole thermodynamics, apply it to the solutions obtained in Sec. \ref{sec:solutions}, and interpret the van der Waals-like critical behavior in terms of topological defect curves. We summarize our results in Sec. \ref{sec:conclusion}.

\section{Einstein Field Equations for a Kalb--Ramond Background}\label{sec:field-equations}
Motivated by Lorentz-symmetry-breaking scenarios in the SME, we consider the Einstein--Hilbert action nonminimally coupled to a self-interacting Kalb--Ramond two-form field \cite{ABK10},
\begin{equation}
\begin{aligned}
S = & \int d^4 x \sqrt{-g} \biggl[ \frac{1}{2\kappa} (R - 2\Lambda) - \frac{1}{12} H^{\mu\nu\rho} H_{\mu\nu\rho}
- V(B^{\mu\nu} B_{\mu\nu}) + \frac{1}{2\kappa} (\xi_1 B^{\mu\nu} B_{\mu\nu} R \\
& + \xi_2 B^{\rho\mu} B^\nu{}_\mu R_{\rho\nu}) \biggr] +\frac{1}{2\kappa} \int d^4 x \sqrt{-g} \mathcal{L}_M,
\end{aligned}
\end{equation}
where we have set $G=1,c=1$ for convenience, and $\Lambda$ is the cosmological constant. The two nonminimal couplings retained from the SME gravitational sector are controlled by $\xi_1$ and $\xi_2$. The matter Lagrangian is chosen as $\mathcal{L}_M=-\frac{1}{2}F^{\mu\nu}F_{\mu\nu}-\eta B^{\alpha\beta}B^{\gamma\rho}F_{\alpha\beta}F_{\gamma\rho}$ \cite{DZY23}, where $\eta$ is a coupling constant and $F_{\mu\nu}=\partial_{\mu}A_{\nu}-\partial_{\nu}A_{\mu}$ is the electromagnetic field strength. The normalization of the Maxwell term is kept fixed throughout the paper, so the charge parameter below should be understood in this
convention.

The Kalb–Ramond field is an antisymmetric rank-two tensor whose kinetic field strength is gauge invariant. Its field strength is defined by \cite{HY01}
\begin{equation}
 H_{\mu\nu\rho} = \partial_{\mu} B_{\nu\rho} + \partial_{\rho} B_{\mu\nu} + \partial_{\nu} B_{\rho\mu},   
\end{equation}
and is invariant under the two-form gauge transformation
\begin{equation}
    B_{\nu\rho} \to B_{\nu\rho} + \partial_{\nu} \Lambda_{\rho} - \partial_{\rho} \Lambda_{\nu},
\end{equation}
where $\Lambda_{\mu}$ is an arbitrary vector field.

To generate a nonzero vacuum expectation value (VEV), $\langle B_{\mu\nu}\rangle=b_{\mu\nu}$, we introduce a self-interaction potential for the Kalb--Ramond field \cite{MCAA19}. A general form is
\begin{equation}
     V=V(B^{\mu\nu}B_{\mu\nu}\pm b^2),
\end{equation}
where $b^{\mu\nu}b_{\mu\nu}=\mp b^2$ and the sign is chosen so that $b^2>0$. The VEV is fixed by the vacuum condition $V(B^{\mu\nu}B_{\mu\nu}\pm b^2)=0$, with the potential minimized at $B^{\mu\nu}B_{\mu\nu}\pm b^2=0$.

It is convenient to decompose the antisymmetric tensor $B_{\mu\nu}$ as
\begin{equation}
 B_{\mu\nu} = \tilde{E}_{[\mu} v_{\nu]} + \epsilon_{\mu\nu\alpha\beta} v^{\alpha} \tilde{B}^{\beta}.
\end{equation}
In spherical coordinates $(t,r,\theta,\phi)$, we take the Kalb--Ramond field configuration to be \cite{LSMA20}

\begin{equation}
B_{\mu\nu} = \begin{pmatrix}
0 & -\tilde{E} & 0 & 0 \\
\tilde{E} & 0 & 0 & 0 \\
0 & 0 & 0 & 0 \\
0 & 0 & 0 & 0
\end{pmatrix},
\end{equation}
For the static radial ansatz used below, the corresponding Kalb--Ramond field strength vanishes identically, $H_{\mu\nu\rho}=0$.

To construct electrically charged black holes, we choose the electrostatic potential $A_{\mu}=-\Phi(r)\delta^t_{\mu}$. A nontrivial electromagnetic coupling to the Kalb--Ramond background is required for a consistent charged solution. Although one may modify the Kalb--Ramond field strength by adding Yang--Mills and gravitational Chern--Simons three-forms, for example $\tilde{H}_{\mu\nu\rho}=H_{\mu\nu\rho}+A_{[\mu}F_{\nu\rho]}$ \cite{MS99}, the corresponding interactions in the kinetic term vanish for the present ansatz \cite{DZY23}. We therefore adopt the matter Lagrangian specified above, which captures the leading algebraic coupling between the electric field and the fixed Kalb--Ramond background.

Reference \cite{LWWL25} showed that the term $\xi_1 B^{\mu\nu}B_{\mu\nu}R$ cannot, in general, be absorbed into a redefinition of the Einstein--Hilbert term in vacuum, because its metric variation contributes nontrivially to the field equations. We therefore retain both nonminimal curvature couplings throughout the analysis.

The metric variation gives
\begin{equation}
    G_{\mu\nu} + \Lambda g_{\mu\nu} = \kappa(T_{\mu\nu}^{M} + T_{\mu\nu}^{EM}),
\end{equation}
with
\begin{equation}
\begin{aligned}
T_{\mu \nu}^{\mathrm{M}} = & \frac{1}{2} H_{\mu \alpha \beta} H_{\nu}^{\alpha \beta}-\frac{1}{12} g_{\mu \nu} H^{\alpha \beta \rho} H_{\alpha \beta \rho} \\
& +\frac{\xi_{1}}{\kappa}\left[\nabla_{\mu} \nabla_{\nu}\left(B^{\alpha \beta} B_{\alpha \beta}\right)-g_{\mu \nu} \nabla^{2}\left(B^{\alpha \beta} B_{\alpha \beta}\right)-B^{\alpha \beta} B_{\alpha \beta} G_{\mu \nu}+2 B_{\mu}^{\alpha} B_{\nu \alpha} R\right] \\
& +\frac{\xi_{2}}{\kappa}\Big[\frac{1}{2} g_{\mu \nu} B^{\alpha \gamma} B_{\gamma}^{\beta} R_{\alpha \beta}-B_{\mu}^{\alpha} B_{\nu}^{\beta} R_{\alpha \beta}-B^{\alpha \beta} B_{\nu \beta} R_{\mu \alpha}-B^{\alpha \beta} B_{\mu \beta} R_{\nu \alpha} \\
& +\frac{1}{2} \nabla_{\alpha} \nabla_{\mu}\left(B^{\alpha \beta} B_{\nu \beta}\right)+\frac{1}{2} \nabla_{\alpha} \nabla_{\nu}\left(B^{\alpha \beta} B_{\mu \beta}\right)-\frac{1}{2} \nabla^{\alpha} \nabla_{\alpha}\left(B_{\mu}^{\gamma} B_{\nu \gamma}\right) \\
& -\frac{1}{2} g_{\mu \nu} \nabla_{\alpha} \nabla_{\beta}\left(B^{\alpha \gamma} B_{\gamma}^{\beta}\right)\Big]+4 V^{\prime}(X) B_{\alpha \mu} B_{\nu}^{\alpha}-g_{\mu \nu} V(X).
\end{aligned}
\end{equation}
The electromagnetic energy-momentum tensor is
\begin{equation}
    T_{\mu\nu}^{EM} =  2F_{\mu\alpha}F_{\nu}^{\alpha} - \frac{1}{2}g_{\mu\nu}F^{\alpha\beta}F_{\alpha\beta} + \eta\left( B^{\alpha\beta}B_{\nu}^{\gamma}F_{\alpha\beta}F_{\mu\gamma} - g_{\mu\nu}B^{\alpha\beta}B^{\gamma\rho}F_{\alpha\beta}F_{\gamma\rho} \right) .
\end{equation}

Varying the action with respect to the vector potential $A^\mu$ gives the modified Maxwell equation
\begin{equation}
     \nabla^\nu \left( F_{\mu\nu} + 2\eta B_{\mu\nu} B^{\alpha\beta} F_{\alpha\beta} \right) = 0.
     \label{Maxwell}
\end{equation}

In the following, the Kalb--Ramond field is treated as a fixed external tensor background, its independent dynamical equation is not analyzed. This assumption is part of the effective-background approximation used in the present work. We now solve the resulting gravitational field equations.

\section{Static Spherically Symmetric Charged Kalb--Ramond Black Hole Solutions}\label{sec:solutions}

We take the static, spherically symmetric metric ansatz
\begin{equation}
    ds^2=-A(r)dt^2+S(r)dr^2+r^2d\theta^2+r^2\sin^2\theta d\phi^2,
\end{equation}
which describes a $(3+1)$-dimensional static and spherically symmetric spacetime.

When the Kalb--Ramond field acquires a nonzero VEV satisfying $g^{\mu\alpha}g^{\nu\beta}b_{\mu\nu}b_{\alpha\beta}=-b^2$, the pseudo-electric component can be written as
 \begin{equation}
     \tilde{E}(r)=|b|\sqrt{\frac{A(r)S(r)}{2}}.
 \end{equation}

For a purely electrically charged black hole, the electrostatic vector potential is $A_{\mu}=-\Phi(r)\delta^t_{\mu}$, and the nonvanishing components of the field strength are
\begin{equation}
     F_{tr}=-F_{rt}=\Phi'(r).
\end{equation}

Substituting the metric ansatz and the VEV configuration into the field equations gives
\begin{equation}
\begin{aligned}
&\frac{2l_{1}+l_{2}}{4S(r)}\left(\frac{A'(r)^{2}}{A(r)}-2A''(r)\right)-\frac{4l_{1}+l_{2}}{2rS(r)}A'(r) +\frac{S'(r)}{S(r)^{2}}\left(\frac{2l_{1}+l_{2}}{4}A'(r)+\frac{A(r)(1+l_{1})}{r}\right) \\
&\quad +\frac{A(r)}{r^{2}S(r)}\left((1+l_{1})(S(r)-1)+\frac{l_{2}}{2}\right) -A(r)\left(V+2b^{2}V'+ \Lambda\right)-\frac{1-6b^{2}\eta}{S(r)}\Phi'(r)^{2}=0,
\end{aligned}
\label{eq1}
\end{equation}

\begin{equation}
\begin{aligned}
&\frac{2l_{1}+l_{2}}{2}\frac{A''(r)}{A(r)}-\frac{2l_{1}+l_{2}}{4}\left(\frac{A'(r)}{A(r)}\right)^{2}+\frac{(1+l_{1})A'(r)}{rA(r)}-\frac{2l_{1}+l_{2}}{4}\frac{A'(r)S'(r)}{A(r)S(r)}-\frac{4l_{1}+l_{2}}{2}\frac{S'(r)}{rS(r)} \\
&\quad +\frac{(1+l_{1})(1-S(r))}{r^{2}}-\frac{l_{2}}{2r^{2}}+S(r)\left(V+2b^{2}V'+\Lambda\right)+\frac{1-6b^{2}\eta}{A(r)}\Phi'(r)^{2}=0,
\end{aligned}
\label{eq2}
\end{equation}

\begin{equation}
\begin{aligned}
&\frac{(2l_{1}+l_{2}-2)r^{2}A'(r)S'(r)}{8A(r)S(r)^{2}}+\frac{(2l_{1}+l_{2}-2)r^{2}A'(r)^{2}}{8A(r)^{2}S(r)}-\frac{(2l_{1}+l_{2}-2)r(rA''(r)+A'(r))}{4A(r)S(r)} \\
&\quad +\frac{(2l_{1}+l_{2}-2)rS'(r)}{4S(r)^{2}}+\Lambda r^{2}+Vr^{2}+4b^{2}V'r^{2}-\frac{r^{2}(1-2b^{2}\eta)\Phi'(r)^{2}}{A(r)S(r)}=0,
\end{aligned}
\label{eq3}
\end{equation}
where we have introduced the dimensionless Lorentz-violating parameters $l_1=b^2\xi_1$ and $l_2=b^2\xi_2$. The factor $\kappa=8\pi$ is absorbed into the definition of the charge parameter $Q$. These parameters may in principle take either sign, but they are assumed to be small. The representative bound used below for a generic dimensionless Lorentz-violating parameter $l$ is \cite{PhysRevD.108.124004}
\begin{equation}
     -1.1 \times 10^{-10} \le l \le 5.4 \times 10^{-10}.
\end{equation}

We next solve these equations for several choices of the potential and cosmological constant.

\subsection{Case A: $V(X)=\frac{\lambda}{2}X^2$, $\Lambda=0$}

We first set the cosmological constant to zero and construct a Reissner--Nordstr\"om-like solution. The vacuum conditions $V=0$ and $V'=0$ place the VEV at a local minimum of the potential. For definiteness, we choose the smooth quadratic potential
\begin{equation}
    V(B^{\mu\nu}B_{\mu\nu}\pm b^2)=V(X)=\frac{\lambda}{2}X^2,
\end{equation}
where $\lambda$ is a constant. The acquisition of a VEV by the Kalb--Ramond field is analogous to spontaneous symmetry breaking in the Higgs mechanism, and a corresponding mass-generation mechanism can be formulated \cite{PhysRevD.40.1886}. In the present background treatment, only the vacuum value and its induced curvature couplings enter the black hole solution.

Combining Eqs. \eqref{eq1} and \eqref{eq2} gives
\begin{equation}
    S(r)=\frac{1}{A(r)}.
\end{equation}
Substitution into the modified Maxwell equation \eqref{Maxwell} yields
\begin{equation}
    \Phi ''+\frac{2}{r}\Phi'=0.
\end{equation}
The electrostatic potential is therefore
\begin{equation}
    \Phi=\frac{c_1}{r}.
    \label{eqPhi}
\end{equation}
The modified Maxwell equation implies the conserved electric flux. By Stokes' theorem, the corresponding electric charge is \cite{DZY23}
\begin{equation}
\begin{aligned}
Q &= -\frac{1}{4\pi} \int_{\Sigma} d^3x \sqrt{\gamma^{(3)}} n_{\mu} J^{\mu} = -\frac{1}{4\pi} \int_{\partial\Sigma} d\theta d\phi \sqrt{\gamma^{(2)}} n_{\mu} \sigma_{\nu} \left( F^{\mu\nu} + 2\eta B^{\mu\nu} B^{\alpha\beta} F_{\alpha\beta} \right) \\
&= \left( 1 - 2b^2\eta \right) c_1,
\end{aligned}
\end{equation}
where $\Sigma$ is a $(2+1)$-dimensional hypersurface with induced metric $\gamma_{ij}^{(3)}$. The integration constant is then fixed as $c_1=Q/(1-2b^2\eta)$, and Eq. \eqref{eqPhi} becomes
\begin{equation}
    \Phi(r)=\frac{Q}{(1-2b^2\eta)r}.
\end{equation}

Substituting $V(X)$ and $\Phi(r)$ into the field equations yields
\begin{equation}
\begin{aligned}
   & A(r)=\frac{1+l_1}{1+l_1-\frac{l_2}{2}}-\frac{2M}{r}+\frac{4(1-l_1)Q^2}{(2-2l_1-l_2)^2r^2},\\
   & S(r)=\frac{1}{A(r)},\\
   & \tilde{E}(r)=\frac{\sqrt{2}}{2}|b|.
\end{aligned}
\label{duguiA}
\end{equation}
The coupling constant $\eta$ is constrained by
\begin{equation}
    \eta =\frac{l_2}{4b^2(1-l_1)}.
\end{equation}
The corresponding horizon radii are
\begin{equation}
r_{\pm}=\frac{\left(1+l_1-\frac{l_2}{2}\right)}{(1+l_1)}\left(M\pm \sqrt{M^2-\frac{(1-l_1^2)\,Q^2}{\left(1+l_1-\frac{l_2}{2}\right)\left(1-l_1-\frac{l_2}{2}\right)^2}}\right)    
\end{equation}

Defining $\gamma \equiv l_2/(2+2l_1),$ the metric function can be rewritten as
\begin{equation}
A(r)=\frac{1}{1-\gamma}-\frac{2M}{r}+\frac{4(1-l_1)Q^2}{(2-2l_1-l_2)^2r^2}.
\end{equation}
It is evident that $A(r)\to 1/(1-\gamma)$ and $S(r)\to 1-\gamma$ at infinity. The electric charge contributes only through terms falling off as $1/r$ and $1/r^2$,  so it does not modify the leading asymptotic structure. The Kretschmann scalar of the charged solution is
\begin{equation}
\begin{aligned}
K=R_{\alpha\beta\delta\lambda}R^{\alpha\beta\delta\lambda}=&\frac{4\gamma^2}{(1-\gamma)^2\,r^4}
-\frac{16\gamma}{1-\gamma}\frac{M}{r^5}
+\frac{48M^2}{r^6}+\frac{8\gamma(1-l_1)}{\left(1-\gamma\right)\left(1-l_1-\frac{l_2}{2}\right)^2}\frac{Q^2}{r^6} \\
&
-\frac{96(1-l_1)}{\left(1-l_1-\frac{l_2}{2}\right)^2}\frac{MQ^2}{r^7}
+\frac{56(1-l_1)^2}{\left(1-l_1-\frac{l_2}{2}\right)^4}\frac{Q^4}{r^8}.
\end{aligned}
\end{equation}
For $Q=0$ this reduces to the neutral result reported in Ref. \cite{LWWL25},  while for $\gamma\to0$ one recovers the Reissner--Nordstr\"om expression $K=48M^2/r^6-96MQ^2/r^7+56Q^4/r^8$.  Since $K$ depends nontrivially on the Lorentz-violating parameters, these effects are intrinsic to the theory and cannot be removed by any coordinate transformation. By performing the coordinate transformations
\begin{equation}
    dt=\sqrt{1-\gamma}\,d\hat{t}, \qquad dr=\sqrt{1/(1-\gamma)}\,d\hat{r},
\end{equation}
the asymptotic metric becomes
\begin{equation}
ds^2=-d\hat{t}^2+d\hat{r}^2+\frac{1}{1-\gamma}\,\hat{r}^2d\Omega^2\,.
\end{equation}
This shows that the temporal and radial sectors coincide with those of Minkowski spacetime in spherical coordinates, while the angular part acquires a constant factor $1/(1-\gamma)$. Because the Maxwell energy-momentum tensor is traceless in four dimensions, the electric charge does not contribute to the Ricci scalar, which evaluates to
\begin{equation}
R=-\frac{2\gamma}{(1-\gamma)r^2}\,.
\end{equation}
Hence, the spacetime is not asymptotically Minkowski. From another perspective, the asymptotic form of the metric in the original coordinates can be written as
\begin{equation}
ds^2=-d\hat{t}^2+(1-\gamma)\,dr^2+r^2d\Omega^2\,,
\end{equation}
so that the deviation from Minkowski spacetime in the radial component is exactly the constant $\gamma$, rather than decaying as $\mathcal{O}(r^{-1})$. This shows that the charged Case~A solution is not asymptotically flat.

\subsection{Case B: $V(X)=\frac{\lambda}{2}X^2$, $\Lambda\ne0$}

We next keep the quadratic potential but allow for a nonzero cosmological constant. The solution is evaluated on the vacuum branch $X=0$, so that $V=V'=0$ for the quadratic potential. The electrostatic potential keeps the same form as in Case A. For $l_2=-4l_1$, equivalently $\xi_2=-4\xi_1$, the field equations admit the exact solution
\begin{equation}
\begin{aligned}
    & A(r)=\frac{(1+l_1)}{(1+3l_1)} - \frac{2M}{r} - \frac{\Lambda}{3(1+l_1)} r^2 + \frac{(1-l_1)Q^2}{(1+l_1)^2 r^2},\\
    & S(r)=\frac{1}{A(r)},\\
    & \tilde{E}(r)=\frac{\sqrt{2}}{2}|b|.
\end{aligned}
\label{dugui}
\end{equation}
The coupling constant $\eta$ is then fixed by
\begin{equation}
    \eta =\frac{-l_1}{b^2(1-l_1)}.
\end{equation}
In this case, the constant $\lambda$ does not need to account for the dependence on the cosmological constant.

\subsection{Case C: $V(X)=\frac{\lambda}{2}X$, $\Lambda\ne0$}
Finally, keeping the cosmological constant nonzero, we choose a linear potential,
\begin{equation}
    V(\lambda,X)=\frac{\lambda}{2}X,
    \label{eqlam}
\end{equation}
where $\lambda$ is a Lagrange multiplier field. In contrast to Cases A and B, $V'\neq0$ when $\lambda$ is nonzero. Substituting Eq. \eqref{eqlam} into Eqs. \eqref{eq1}--\eqref{eq3}, one obtains an exact analytical solution provided that the coupling constants satisfy
\begin{equation}
    \lambda=\frac{(4\xi_1+\xi_2)\Lambda}{(1-l_1-\frac{l_2}{2})},\qquad \eta =\frac{l_2}{4b^2(1-l_1)}.
\end{equation}
In case C, the Lagrange multiplier field $\lambda$ depends on the cosmological constant. This condition follows directly from the field equations and is required for the closed-form solution displayed below. The resulting solution is
\begin{equation}
\begin{aligned}
   & A(r)=\frac{1+l_1}{1+l_1-\frac{l_2}{2}}-\frac{2M}{r}+\frac{4(1+l_1)Q^2}{(2-2l_1-l_2)^2r^2}- \frac{\Lambda}{3(1-l_1-\frac{l_2}{2})} r^2,\\
   & S(r)=\frac{1}{A(r)},\\
   & \tilde{E}(r)=\frac{\sqrt{2}}{2}|b|.
\end{aligned}
\label{duguiC}
\end{equation}

\section{Thermodynamic Analysis and the Joule--Thomson Expansion}\label{sec:thermodynamics}

\subsection{Iyer--Wald Formalism}\label{sec:iyer-wald}

Black hole thermodynamics exposes the interplay among gravity, thermodynamics, and quantum theory. In Einstein gravity, the Bekenstein--Hawking area law provides the entropy, while the Hawking temperature is determined by the surface gravity. In theories with nonminimal curvature couplings, however, the entropy need not be given solely by the horizon area. 

The Iyer--Wald covariant phase-space formalism is therefore the appropriate tool: it constructs the presymplectic potential and Noether charge covariantly, identifies the black hole entropy with the Noether charge evaluated on the bifurcation surface, and yields the first law and Smarr relation \cite{r1,r2,by8,by18,by19,by20}. In the present model, the two Kalb--Ramond curvature couplings correct both the energy and the entropy, and these corrections must be retained consistently before one discusses phase behavior. In this chapter, we first discuss the thermodynamic functions of the black holes in Cases A and B, and then investigate the Joule--Thomson expansion of the Case B black hole.

Consider the $（3+1）$-dimensional spacetime Lagrangian form for the gravitational, Kalb--Ramond, and electromagnetic sectors,
\begin{equation}
    \mathbf{L}  = \mathcal{L}\epsilon,
\end{equation}
where $\mathcal{L}$ is the total Lagrangian density and $\epsilon$ is the spacetime volume form. Varying the Lagrangian with respect to the fields $\Phi=\{g_{ab},A_a\}$ gives
\begin{equation}
    \delta \mathbf{L} = \mathbf{E}[\Phi] \delta \Phi + d \boldsymbol{\Theta}[\Phi, \delta \Phi].
    \label{45}
\end{equation}
After isolating the bulk equations of motion, the presymplectic potential of the system can be written as
\begin{equation}
\begin{matrix}
\Theta[\Phi,\delta\Phi]_{\mu\nu\rho} = &\Big[2E_R{}^{\lambda\sigma\tau\kappa}\nabla_{\kappa}\delta g_{\sigma\tau} - 2\big(\nabla_{\kappa}E_R{}^{\lambda\sigma\tau\kappa}\big)\delta g_{\sigma\tau} \\
&+ \big(-2F^{\lambda\tau} - 4\eta B^{\alpha\beta}B^{\lambda\tau}F_{\alpha\beta}\big)\delta A_{\tau}\Big]\epsilon_{\lambda\mu\nu\rho},
\label{eq:4.3}
\end{matrix} \\
\end{equation}
where
\begin{equation}
    E_R{}^{\lambda\sigma\tau\kappa} = \frac{1}{2k}\Big(X^{\lambda\sigma\tau\kappa} + \xi_2 B^{\mu\rho}B_{\rho}^{\nu}Y_{\mu\nu}{}^{\lambda\sigma\tau\kappa} + \xi_1 B^{\mu\nu}B_{\mu\nu}X^{\lambda\sigma\tau\kappa}\Big),
\label{eq:4.4}
\end{equation}
\begin{equation}
    X^{\lambda\sigma\tau\kappa} = g^{\lambda[\tau}g^{\kappa]\sigma},
\label{eq:4.5}
\end{equation}
\begin{equation}
    Y_{\mu\nu}{}^{\lambda\sigma\tau\kappa} = \frac{1}{2}\Big(g_{(\mu}{}^{\lambda}g_{\nu)}{}^{[\tau}g^{\kappa]\sigma} - g_{(\mu}{}^{\sigma}g_{\nu)}{}^{[\tau}g^{\kappa]\lambda}\Big).
\label{eq:4.6}
\end{equation}
The Lagrangian density is a $D$-form, and its variation under $\delta_{\xi}\Phi=\mathcal{L}_{\xi}\Phi$ is given by
\begin{equation}\label{eq:48}
\delta_{\xi}\mathbf{L}=\mathrm{d}(\xi\cdot\mathbf{L})=\mathbf{E}[\Phi]\mathcal{L}_{\xi}\Phi+\mathrm{d}\mathbf{\Theta}[\Phi,\mathcal{L}_{\xi}\Phi].
\end{equation}
By Noether's second theorem, the first term on the right is exact and vanishes on-shell. The conserved Noether current is
\begin{equation}\label{eqJ}
\mathbf{J}_{\xi}=\mathbf{\Theta}[\Phi,\mathcal{L}_{\xi}\Phi]-\xi\cdot\mathbf{L},
\end{equation}
which is conserved on shell, $d\mathbf{J}_\xi=0$. Hence, by the Poincare lemma, this implies that at least locally there exists a $2$-form Noether charge $\mathbf{Q}_{\xi}$, such that
\begin{equation}
    \mathbf{J}_\xi=d\mathbf{Q}_{\xi} .
\end{equation}
 For the charged Kalb--Ramond black holes, the charge takes the form
\begin{equation}
\begin{matrix}
(\mathbf{Q}_{\xi})_{\mu\nu} = \Big[-E_R{}^{\lambda\sigma\tau\kappa}\nabla_{\tau}\xi_{\kappa} - 2\xi_{\tau}\nabla_{\kappa}E_R{}^{\lambda\sigma\tau\kappa} + \big(-2F^{\lambda\sigma} - 4\eta B^{\alpha\beta}B^{\lambda\sigma}F_{\alpha\beta}\big)A_{\kappa}\xi^{\kappa}\Big]\epsilon_{\lambda\sigma\mu\nu},
\label{eq:4.7}
\end{matrix} \\ 
\end{equation}
Varying Eq. \eqref{eqJ} gives
\begin{equation}
    \mathrm{d} \left( \delta \mathbf{Q}_{\xi} - \xi \cdot \mathbf{\Theta} [\Phi, \delta \Phi] \right) = \delta [\mathbf{\Theta} (\Phi, \mathcal{L}_{\xi} \Phi)] - \mathcal{L}_{\xi} [\mathbf{\Theta} [\Phi, \delta \Phi]].
    \label{eqQ}
\end{equation}
The surface charge associated with the diffeomorphism variation $\delta_{\xi}\Phi=\mathcal{L}_{\xi}\Phi$ is
\begin{equation}
    \delta H_{\xi} = \int_{S_{\infty}} \left(\delta \mathbf{Q}_{\xi} - \xi \cdot \mathbf{\Theta}[\Phi, \delta \Phi]\right).
\end{equation}
For the timelike Killing vector $\partial_t$, the energy variation is
\begin{equation}
    \delta E=\left(1-l_{1}-\frac{l_2}{2}\right)\delta M.
    \label{eqE}
\end{equation}
Since the Killing vector $\xi_{H}$ vanishes on the bifurcation surface $S_h$, integrating Eq. \eqref{eqQ} yields that the Wald entropy satisfies
\begin{equation}
    \int_{S_h} \delta \mathbf{Q}_{\xi_H} = T \delta S_W + \Phi \delta Q.
\end{equation}
Then one can obtain the expression for the Wald entropy
\begin{equation}
    S_{W} = - 2\pi\int_{S_{h}}^{}\tilde{\epsilon}E_{R}{}^{\mu\nu\rho\sigma}\epsilon_{\rho\sigma}\epsilon_{\mu\nu},
\end{equation}
where $\epsilon_{\mu\nu}$ denotes the binormal associated with the bifurcation surface, and $\tilde{\epsilon}$ is a 2-form (i.e., the area element) on the $S_h$.

For Case A, with the metric functions given in Eq. \eqref{duguiA} , the Wald entropy and the energy are
\begin{equation}
    S_{W}=\left(1 - l_{1}- \frac{l_2}{2}\right)\pi r_{h}^{2},
    \label{eqS}
\qquad
    E_W=\left(1 - l_{1}- \frac{l_2}{2}\right)M,
\end{equation}
The corresponding Hawking temperature follows from the surface gravity,
\begin{equation} 
	T = \frac{\kappa}{2\pi} = \frac{1}{4\pi} \left( \frac{2M}{r_h^2} - \frac{8(1-l_1)Q^2}{(2-2l_1-l_2)^2 r_h^3} \right),
\end{equation}
where the event horizon radius $r_h$ is the largest positive real root of $A(r)=0$. Solving the horizon condition for the mass gives
\begin{equation} \label{M}
    M(r_h) = \frac{(1 + l_1) r_h}{2 \left( 1 + l_1 - \frac{l_2}{2} \right)} + \frac{2(1 - l_1) Q^2}{(2 - 2l_1 - l_2)^2 r_h}.
\end{equation}
Within the Iyer--Wald thermodynamic formalism, the thermodynamic quantities corrected by the Lorentz-symmetry-breaking parameters satisfy the Smarr relation
\begin{equation}
    E = 2T_{H}S_{W} + \Phi Q .
\end{equation}
For $P=0$ , we compute the isobaric and isochoric heat capacities of the Case A black hole

\begin{align}
C_{P} &=T{{\left ({\frac{\partial S}{\partial T}}\right )}_{P,Q}} \\
&=2\pi\left(1-l_1-\frac{l_2}{2}\right)r_h^{2}\,
\frac{(1+l_1)(2-2l_1-l_2)^{2}\,r_h^{2}-4(1-l_1)\!\left(1+l_1-\dfrac{l_2}{2}\right)Q^{2}}
{12(1-l_1)\!\left(1+l_1-\dfrac{l_2}{2}\right)Q^{2}-(1+l_1)(2-2l_1-l_2)^{2}\,r_h^{2}},    
\end{align}

\begin{equation}
C_{V}=T{{\left ({\frac{\partial S}{\partial T}}\right )}_{V,Q}}=0.
\end{equation}

For Case B, with the metric given in Eq. \eqref{dugui} and $l_2=-4l_1$, the thermodynamic quantities become
\begin{equation}
    S_{W} = (1 + l_{1})\pi r_{h}^{2}, 
\qquad
     E_W=(1+l_1)M,
\end{equation}
\begin{equation}
    T = \frac{\kappa}{2\pi} = \frac{1}{4\pi} \left( \frac{2M}{r_h^2} - \frac{2(1-{l_1}) Q^2}{(1+{l_1})^2 r_h^3} - \frac{2\Lambda r_h}{3(1+l_1)} \right),
    \label{wendu}
\end{equation}
\begin{equation}
M = \frac{1}{2} r_h \left( \frac{1+l_1}{1+3l_1} + \frac{(1-l_1) Q^2}{(1+l_1)^2 r_h^2} - \frac{\Lambda r_h^2}{3(1+l_1)} \right).
\label{eq:mass}
\end{equation}
In the extended phase space, the cosmological constant is identified with the thermodynamic pressure $P = - {\Lambda}/{(8\pi)}$,and the thermodynamic volume is \cite{Ahmed:2025oxp}
\begin{equation}
V = \left( \frac{\partial E}{\partial P} \right)_{S,Q} = \frac{4\pi}{3} r_h^3.
\label{eq:volume}
\end{equation}
These quantities satisfy the Smarr relation
\begin{equation}
    E = 2T_{H}S_{W} + \Phi Q - 2PV.
\end{equation}
From Eqs. \eqref{eq:mass} and \eqref{wendu}, the equation of state of the charged Kalb--Ramond black hole is

\begin{equation}
P =\frac{(1+l_1)T}{2r_h} + \frac{(1-l_1)Q^2}{8\pi(1+l_1)r_h^4} - \frac{(1+l_1)^2}{8\pi(1+3l_1)r_h^2}.
\label{eq:eos}
\end{equation}
As in van der Waals fluids, this black hole system exhibits critical behavior \cite{Hegde:2024jtgaussbonnet,Kruglov:2022nedadsjt,Kruglov:2022nonlinearchargedjt,Ahmed:2025oxp,Kruglov:2022symmetrynedthermo,Ali:2025nedrnadsjt,Kruglov:2023canadianjpmagnetic,Wang:2025quantumcorrectedadsjt,Cao:2021darkmatterrnadsjt}. The critical point is determined by the inflection-point conditions of $P(r_h,T)$ at fixed temperature \cite{ElHadri:2026noncommutativernads,Mo:2018ddimchargedjt},

\begin{equation}
\frac{\partial P}{\partial r_{h}} = 0,\qquad
\frac{\partial^{2}P}{\partial r_{h}^{2}} = 0,
\end{equation}
which leads to
\begin{equation}
\begin{aligned}
&r_{c} =  \sqrt{6} Q \frac{\sqrt{(1-l_1)(1+3l_1)}}{\sqrt{(1+l_1)^3}},\\
&T_{c} = \frac{(1+l_1)^{5/2}}{3\sqrt{6}\pi Q \sqrt{1-l_1} (1+3l_1)^{3/2}},\\
&P_{c} = \frac{(1+l_1)^{5}}{96(1-l_1)(Q+3l_1Q)^{2}\pi}.
\end{aligned}
\end{equation}
The critical parameters explicitly depend on the Lorentz-violating parameter $l_1$. Other thermodynamic response functions follow from the same relations. For example, the isobaric and isochoric heat capacities are
\begin{equation}
\begin{aligned}
C_{P}&=T{{\left ({\frac{\partial S}{\partial T}}\right )}_{P,Q}}\\&=2(1+l_1)\pi r_{h}^{2} \frac{(1+2l_1-3l_1^2) Q^{2} - (1+l_1)^{3} r_{h}^{2} - 8(1+l_1)(1+3l_1) P \pi r_{h}^{4} }{-3(1+2l_1-3l_1^2) Q^{2} + (1+l_1)^{3} r_{h}^{2} - 8(1+l_1)(1+3l_1) P \pi r_{h}^{4}},
\end{aligned}
\end{equation}
and
\begin{equation}
C_{V}=T{{\left ({\frac{\partial S}{\partial T}}\right )}_{V,Q}}=0.
\end{equation}
We now turn to the Joule--Thomson expansion of these charged Kalb--Ramond black holes.

\subsection{Joule--Thomson Expansion for Case B}\label{sec:joule-thomson}
The Joule--Thomson expansion is defined only for black holes with a nonvanishing cosmological constant, since the thermodynamic pressure is identified as $P=-\Lambda/(8\pi)$ in the extended phase space.And we take Case~B as the representative example.

In a black hole Joule--Thomson expansion, the black hole mass is interpreted as enthalpy and is held fixed. The Joule--Thomson coefficient is therefore defined as $\mu=(\partial T/\partial P)_M$ \cite{Media:2025kerrdslorentzviolationjt,Biswas:2021yangmillsjt,Ahmed:2025oxp,Kruglov:2023gravcosmolmagneticjt,Liang:2021torusblackholejt,Sekhmani:20225drchargedjt,Masmar:2023nonlinearchargedjt,Mo:2018ddimchargedjt,Alipour:2025yangmillskerrsenjt}. Standard thermodynamic identities give

\begin{equation}
\mu = \left( \frac{\partial T}{\partial P} \right)_{M} = \frac{1}{C_{P}}\left\lbrack T\left( \frac{\partial V}{\partial T} \right)_{P} - V \right\rbrack.
\label{eq:jtmu}
\end{equation}
The inversion temperature is obtained by imposing $\mu=0$, which is
\begin{equation}
T_{i} = V\left(\frac{\partial T}{\partial V}\right)_P.
\end{equation}
Using Eq. \eqref{eq:volume}, the equation of state can be expressed in terms of the thermodynamic volume as
\begin{equation}
T =-\frac{(1-l_1)Q^2}{3(1+l_1)^2 V} + \frac{6^{2/3} (1+l_1)}{12 (1+3l_1) \pi^{2/3} V^{1/3} } + \frac{6^{1/3}  P V^{1/3}}{(1+l_1) \pi^{1/3}}.
\label{eq:TV}
\end{equation}
Equation \eqref{eq:TV} then gives the inversion temperature
\begin{equation}
T_{i} = \frac{(1-l_1)Q^2}{3(1+l_1)^2 V} - \frac{6^{2/3}(1+l_1)}{36(1+3l_1)\pi^{2/3} V^{1/3}} + \frac{6^{1/3} P_{i} V^{1/3}}{3(1+l_1)\pi^{1/3}},
\end{equation}
where $P_i$ denotes the inversion pressure.
Substituting $P=P_i$ into Eq. \eqref{eq:TV} yields an equivalent expression
\begin{equation}
T_i=-\frac{(1-l_1)Q^2}{3(1+l_1)^2 V} + \frac{6^{2/3} (1+l_1)}{12 (1+3l_1) \pi^{2/3} V^{1/3} } + \frac{6^{1/3}  P_i V^{1/3}}{(1+l_1) \pi^{1/3}}.
\end{equation}
Equating the two expressions gives
\begin{equation}
(1+l_1)^3 V^{2/3} + 6^{2/3} (1+l_1)(1+3l_1) P_i V^{4/3} \pi^{1/3} - 6^{1/3} (1+3l_1)(1-l_1) Q^2 \pi^{2/3} = 0.
\end{equation}
After substituting $V=4\pi r_h^3/3$, the resulting equation has four roots for $r_h$; the physically relevant one is the positive real root
\begin{equation} 
r_{h} = \frac{1}{2\sqrt{2}}
\sqrt{
  \frac{
    F(l_1,P_i,Q)
  }{
    \pi(1+l_1)(1+3l_1) P_i 
  }
},
\end{equation}
where the function $F$ is given by
\begin{equation}
F(l_1,P_i,Q)=\sqrt{(1+l_1)^6 + 24\pi (1-l_1^2)  (1+3l_1)^2 P_i Q^2 } - (1+l_1)^3.
\end{equation}
Substituting this root into the inversion-temperature formula gives
\begin{equation} \label{Inversion-temperature formula}
T_{i}  = \frac{
    16 (1-l_1) Q^2 \pi 
    - \dfrac{(1+l_1)^2 F(l_1,P_i,Q)}{(1+3l_1)^2 P_i}
}{
    \sqrt{2}\pi^2  (1+l_1)^2 
    \left(
        \dfrac{ F(l_1,P_i,Q) }{\pi(1+l_1)(1+3l_1) P_i }
    \right)^{3/2}
} .
\end{equation} 
In the limit $P_i\to0$, the minimum inversion temperature is
\begin{equation}
T_{i}^{\min} = \dfrac{(1+l_1)^{5/2}}{6\sqrt{6}\pi  \sqrt{1-l_1} (1+3l_1)^{3/2}Q}.
\end{equation}
Consequently,
\begin{equation}
\frac{T_{i}^{\min}}{T_{c}} = \frac{1}{2}.
\end{equation}

Although the Kalb--Ramond background changes the separate values of $T_i^{\min}$ and $T_c$, their ratio remains unchanged. This invariance can be understood geometrically. The parameter $l_1$ acts as a reparametrization of the thermodynamic phase space. In particular, under the rescaling
  \begin{equation}
  \tilde{T} = (1+l_{1})\,T, 
\qquad
 \tilde{v} = \frac{v}{1+l_{1}},
  \end{equation}
the equation of state reduces to the canonical charged-AdS form 
\begin{equation}
    P = \frac{\tilde{T}}{\tilde{v}} - \frac{\tilde{a}}{\tilde{v}^{2}} + \frac{\tilde{b}}{\tilde{v}^{4}},
\end{equation}
with redefined constants $\tilde a$ and $\tilde b$. The dimensionless ratio $T_i^{\min}/T_c$ is invariant under this reparametrization. In this sense, the charged Kalb--Ramond black hole is thermodynamically equivalent, at the level of dimensionless ratios, to the Reissner--Nordstr\"om--AdS black hole within this one-parameter sector.

Fig.\ref{fig:T-P} shows the isenthalps given by Eq. \eqref{eq:mass} and \eqref{eq:eos} and the inversion curve given by Eq. \eqref{Inversion-temperature formula} in the $T$-$P$ plane. The inversion curve, defined by $\mu=0$, separates the $T$-$P$ plane into a cooling region ($\mu>0$), where the temperature decreases during the expansion, and a heating region ($\mu<0$), where it increases. It also marks the locus on which the slope of each isenthalpic curve vanishes.

\FloatBarrier
\begin{figure}[htbp]
\vspace{0.5cm}
    \centering
    \begin{subfigure}[b]{0.49\textwidth}
        \centering
        \adjustbox{width=\linewidth, totalheight=0.28\textheight, keepaspectratio, center}{
            \includegraphics{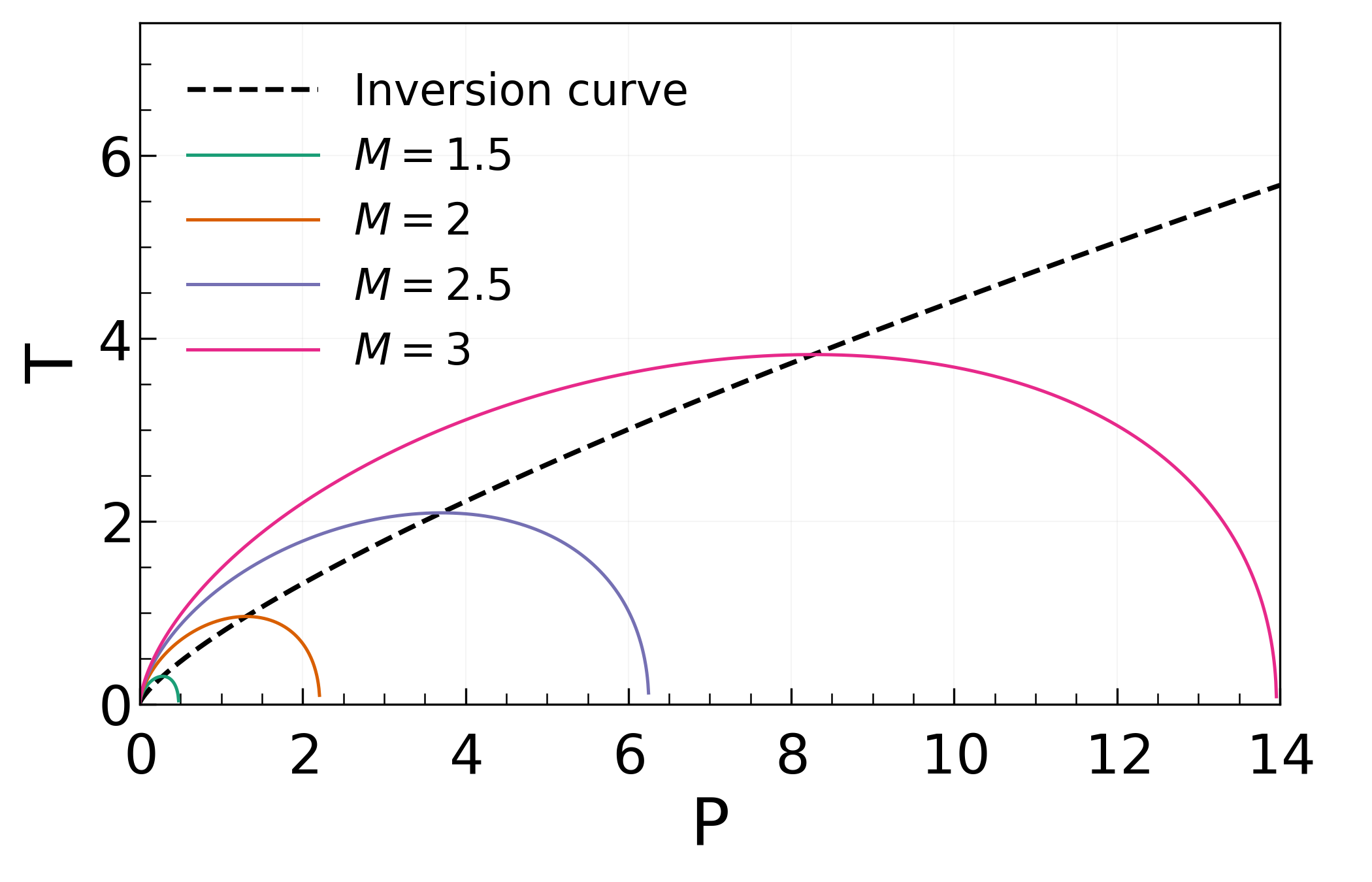}
        }
        \caption{$Q = 1$, $l_1 = 1 \times 10^{-10}$}
        \label{fig:sub1}
    \end{subfigure}
    \hfill
    \begin{subfigure}[b]{0.49\textwidth}
        \centering
        \adjustbox{width=\linewidth, totalheight=0.28\textheight, keepaspectratio, center}{
            \includegraphics{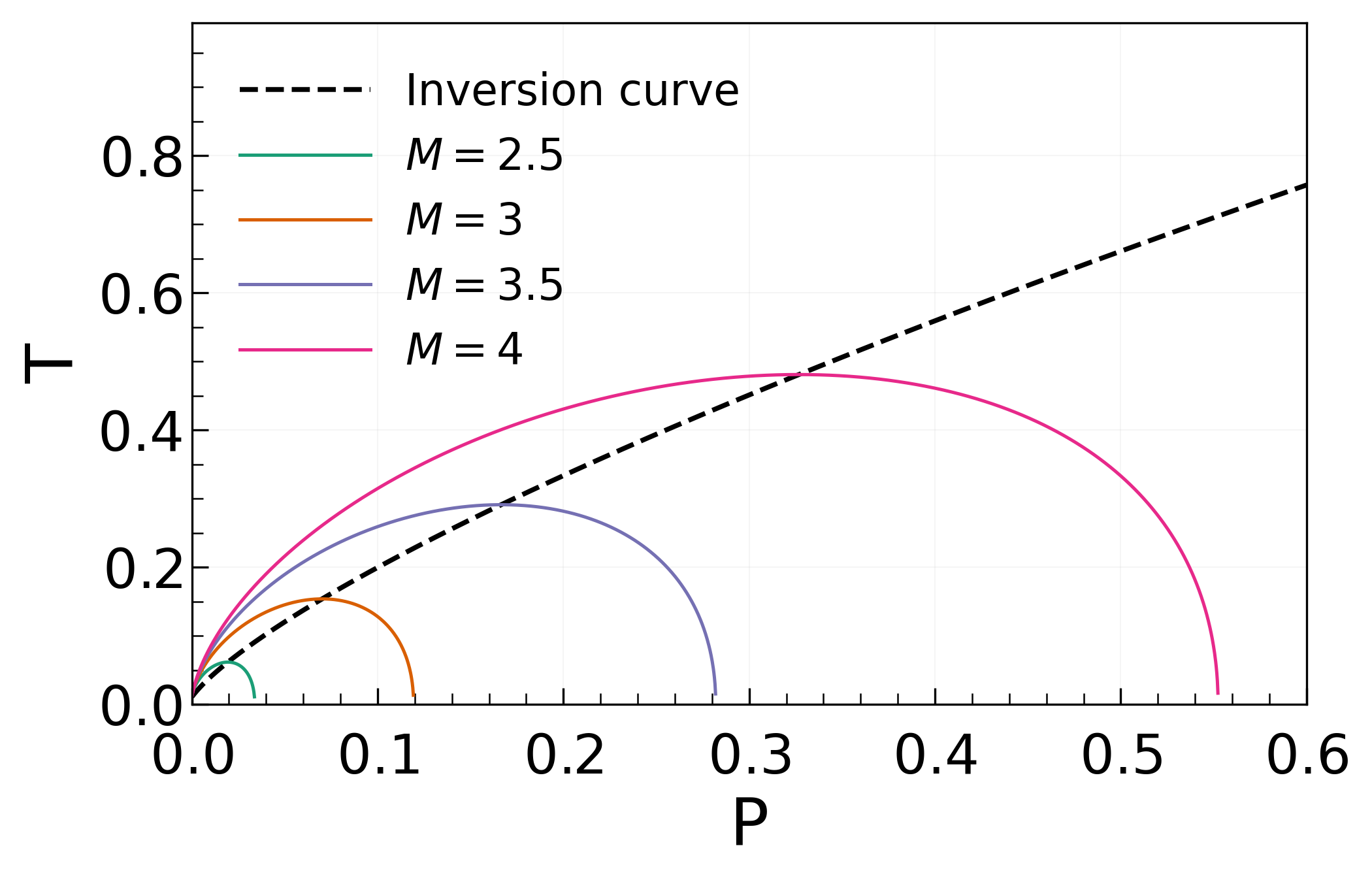}
        }
        \caption{$Q = 2$, $l_1 = 1 \times 10^{-10}$}
        \label{fig:sub2}
    \end{subfigure}
\end{figure}

\FloatBarrier
\begin{figure}[htbp]
    \ContinuedFloat  
    \centering
    \begin{subfigure}[b]{0.49\textwidth}
        \centering
        \adjustbox{width=\linewidth, totalheight=0.28\textheight, keepaspectratio, center}{
            \includegraphics{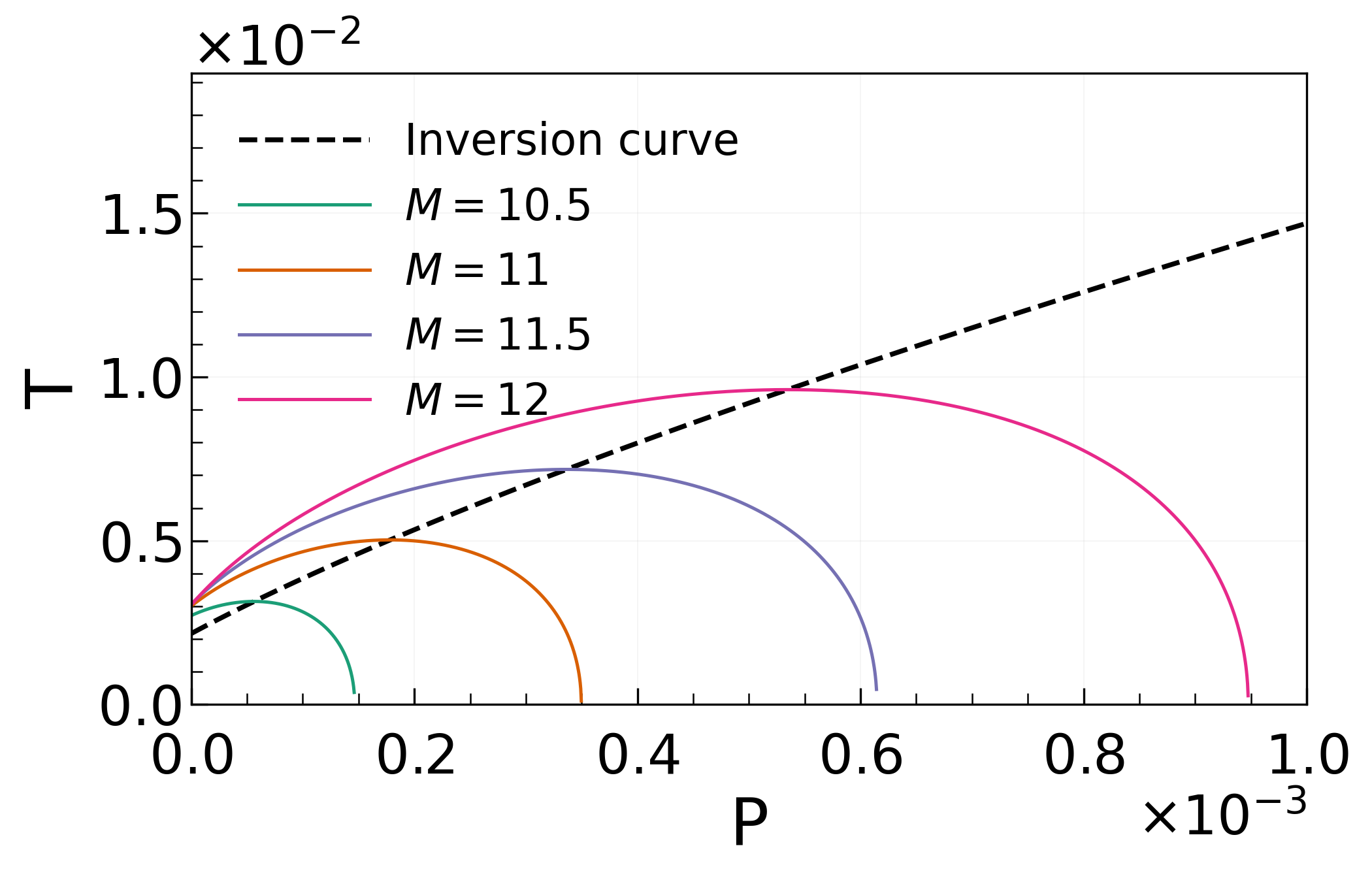}
        }
        \caption{$Q = 10$, $l_1 = 1 \times 10^{-10}$}
        \label{fig:sub3}
    \end{subfigure}
    \hfill
    \begin{subfigure}[b]{0.49\textwidth}
        \centering
        \adjustbox{width=\linewidth, totalheight=0.28\textheight, keepaspectratio, center}{
            \includegraphics{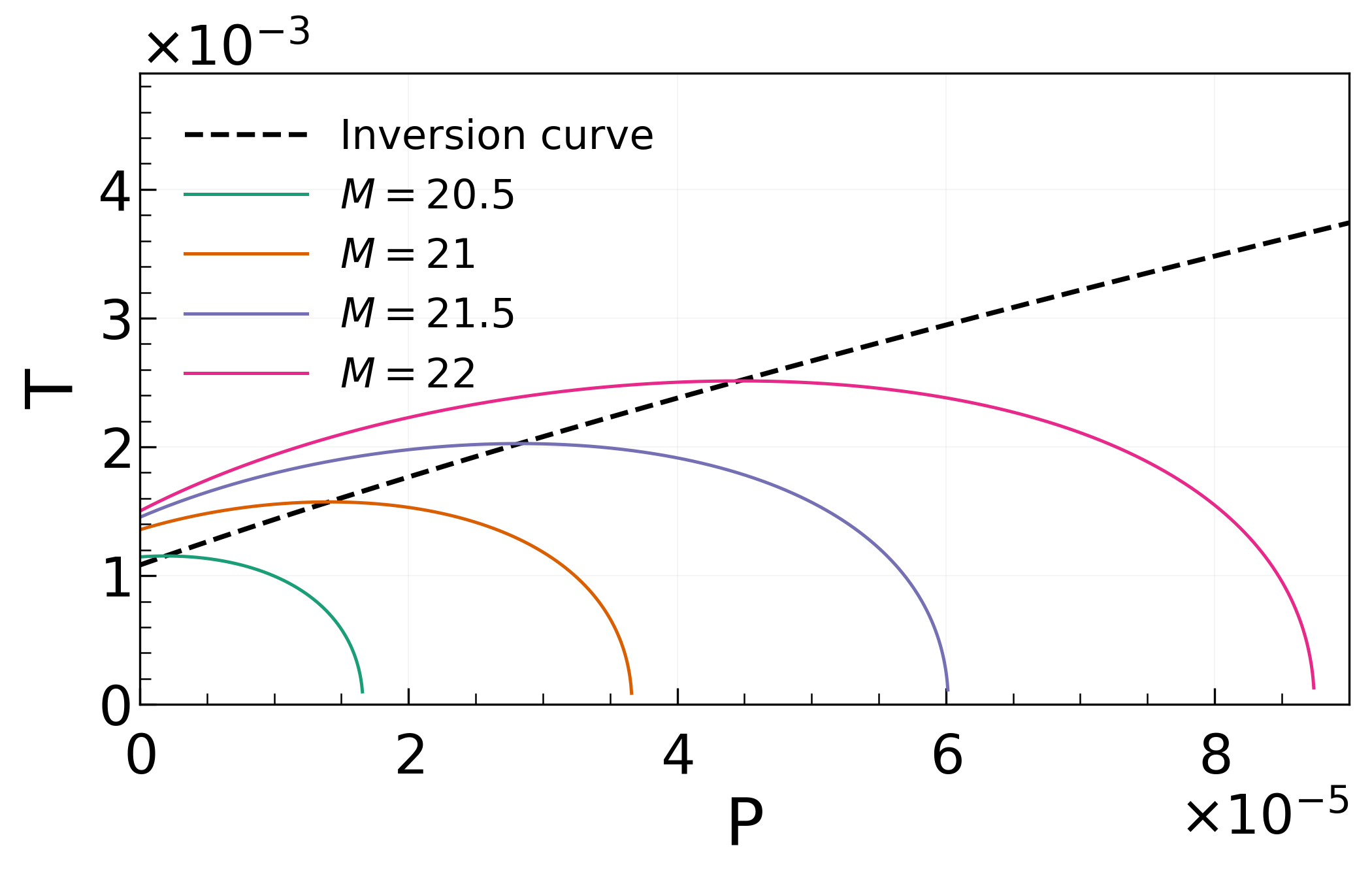}
        }
        \caption{$Q = 20$, $l_1 = 1 \times 10^{-10}$}
        \label{fig:sub4}
    \end{subfigure}
    
    \vspace{0.3cm}
    \caption{Inversion and isenthalpic curves for Case B.
    }
    \label{fig:T-P}
    \vspace{0.2cm}
\end{figure}
\FloatBarrier

\subsubsection{Influence of the Lorentz-Symmetry-Breaking Parameter $l_1$}

\begin{figure}[ht]
    \centering
    \includegraphics[width=0.6\textwidth]{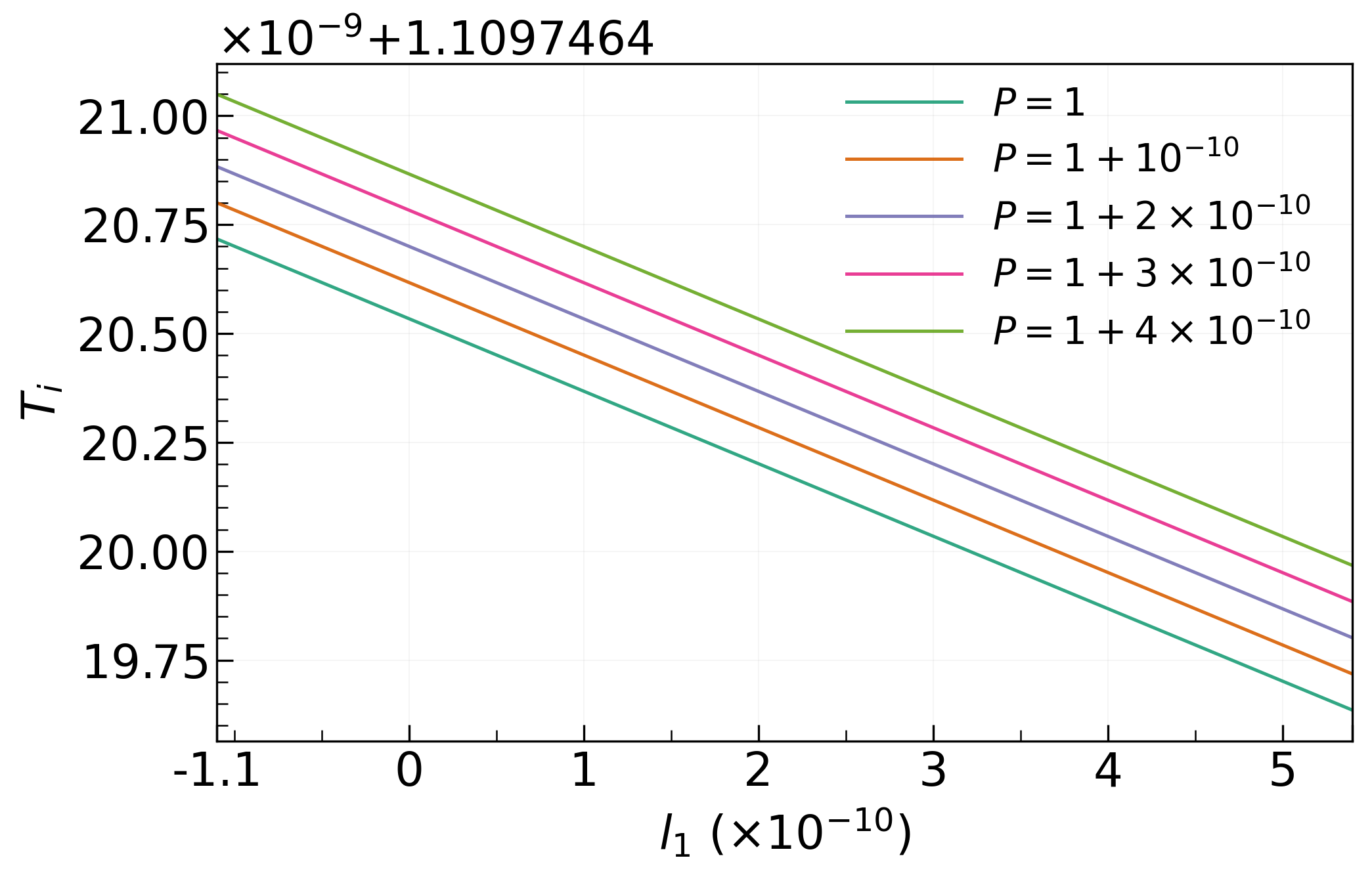}
    \vspace{0.4cm}
    \caption{The inversion temperature $T_i$ as a function of the Lorentz-symmetry-breaking parameter $l_{1}$ for different fixed values of the pressure $P$, with the charge fixed at $Q=2$.}
    \label{fig:l1_inversion}
    \vspace{0.2cm}
  \end{figure}

We further examine how the Lorentz-symmetry-breaking parameter $l_1$ affects the inversion temperature. For fixed charge $Q=2$, Fig. \ref{fig:l1_inversion} shows the inversion temperature as a function of $l_1$ for several fixed pressures near unity.

Within the parameter range considered, the inversion temperature decreases monotonically as $l_1$ increases. The curves for different pressures are nearly parallel, and larger pressures correspond to slightly larger inversion temperatures. Thus $l_1$ systematically shifts the inversion temperature and modifies the relative size of the cooling and heating regions.
\FloatBarrier
\FloatBarrier

\section{Topology and Phase Transitions in Black Hole Thermodynamics}\label{sec:topology}
\subsection{Topological Current}
Phase transitions are a central topic in black hole thermodynamics \cite{a12,a13,a14,a15,a16,a17,a18,a19,a20,a21,a22,a23,a24,a25,a26,a27,a28,a29,a30}. Examples include Davies-type \cite{a12}, Hawking--Page \cite{a13}, extremal \cite{a14,a15,a16,a17,a18,a19,a20,a21,a22}, and van der Waals-type transitions \cite{a23,a24,a25,a26,a27,a28,a29,a30}. A recent development is the topological interpretation of such phase transitions \cite{Wei:2022dzw,Wei:2024gfz}. In this framework, black hole branches are endowed with topological charges, and critical points are interpreted as defects in the thermodynamic parameter space. The construction is based on Duan's $\Phi$-mapping topological current theory \cite{Duan1,Duan2}, which we briefly review below. The topological construction starts from York's generalized off-shell free energy \cite{York}. For a black hole of energy $E$ and entropy $S$ in a cavity with temperature $1/\tau$, it is
\begin{equation}
    \mathcal{F}=E-\frac{S}{\tau}.
\end{equation}

Introducing an auxiliary parameter $\Theta\in(0,\pi)$, one defines the two-component vector field
\begin{equation}
    \phi= \left (\frac{\partial\mathcal{F}}{\partial r_{h}},-\cot\Theta\,\csc\Theta \right ).
\end{equation}
The zero points of $\phi$ satisfy $\Theta=\pi/2$ and $\tau=T^{-1}$. Each zero therefore corresponds to a black hole state \cite{Wei:2024gfz} and can be assigned a topological charge in Duan's theory.

The associated topological current is
\begin{equation}
    j^{\mu}=\frac{1}{2\pi}\epsilon^{\mu\nu\rho}\epsilon_{ab}\partial_{\nu}n^{a} \partial_{\rho}n^{b},
\end{equation}
where $\mu,\nu,\rho=0,1,2$, $\partial_\nu=\partial/\partial x^\nu$, and $x^\nu=(\tau,r_h,\Theta)$. The unit vector is $n^a=\phi^a/\|\phi\|$, with $a=r_h,\Theta$. The current is conserved,
\begin{equation}
    \partial_{\mu}j^{\mu}=0,
\end{equation}
as can be verified directly. Using the Jacobi tensor $\epsilon^{ab}J^\mu(\phi/x)=\epsilon^{\mu\nu\rho}\partial_\nu\phi^a\partial_\rho\phi^b$ and the two-dimensional Green-function identity
\begin{equation}
    \Delta_{\phi^a}\ln\|\phi\|=2\pi\delta^2(\phi),
\end{equation}
the current can be written as \cite{Duan2}
\begin{equation}
    j^{\mu}=\delta^2(\phi)J^{\mu}\left(\frac{\phi}{x} \right),
\end{equation}
which is nonzero only at points satisfying $\phi^a(x^i)=0$. Denoting the $i$-th zero by $\vec{x}=\vec{z}_i$, the density becomes
\begin{equation}
    j^0 = \sum_{i=1}^N \beta_i \eta_i \delta^2 (\vec{x} - \vec{z}_i). 
\end{equation}
In a parameter region $\Sigma$, the corresponding topological number is
\begin{equation}
    W=\int_{\Sigma}^{}j^0d^2x=\sum_{i=1}^{N}\beta_{i}\eta_{i}=\sum_{i=1}^{N}\omega_{i}.
\end{equation}
Here $\beta_i$ is the positive Hopf index, and
\begin{equation}
    \eta_i=\operatorname{sign}[J^0(\phi/x)_{z_i}]=\pm1
\end{equation}
which is the Brouwer degree, and $\omega_i$ is the winding number of the $i$-th zero point. A zero point with $\omega_i=+1$ corresponds to a locally stable branch with positive heat capacity, while $\omega_i=-1$ corresponds to a locally unstable branch with negative heat capacity. We now apply this construction to the charged black holes obtained above and examine how the Lorentz-symmetry-breaking parameters affect their topology.

\subsection{Topological Properties of Charged Kalb--Ramond  Black Holes}

We now analyze the thermodynamic topology of the electrically charged, static, spherically symmetric solutions obtained in Sec. \ref{sec:solutions}. Both the black hole solutions in Case A and Case B can be written in the unified form
\begin{equation}
   ds^{2} = - A(r)dt^{2} + A^{- 1}(r)dr^{2} + r^{2}d\Omega^{2},
\end{equation}

\begin{equation}
   A(r) = c_0 - \frac{2M}r + \frac{ Q^2}{c_1 r^2} - \frac{\Lambda r^2}{3 c_2}  . 
\end{equation}
For Case A,
\begin{equation}
    c_0=\frac{1+l_1}{1+l_1-\frac{l_2}{2}}, c_1=\frac{(2-2l_1-l_2)^2}{4(1-l_1)},  c_2\to\infty.
\end{equation}
and for Case B, 
 \begin{equation}
    c_0=\frac{1+l_1}{1+3l_1}, c_1=\frac{(1+l_1)^2}{1-l_1},  c_2=1+l_1.
\end{equation}
 In both cases, the horizon condition gives
\begin{equation}
    M=\frac{c_0 r_h}{2}+\frac{Q^2}{2 c_1 r_h}-\frac{\Lambda r_h^3}{6 c_2}.
\end{equation}
In Case A, Eqs. \eqref{eqS} gives the energy and entropy, which are
\begin{equation}
E_{W}=\left(1-l_1-\frac{l_2}{2}\right)M,\qquad S_{W}=\left(1-l_1-\frac{l_2}{2}\right)\pi r_h^2.
\end{equation}
The generalized off-shell free energy is therefore
\begin{equation}
    \mathcal{F}=E-\frac{S}{\tau}=\left(1-l_1-\frac{l_2}{2}\right)\left(\frac{c_0 r_h}{2}+\frac{Q^2}{2 c_1 r_h}-\frac{\Lambda r_h^3}{6 c_2}\right)-\frac{\left(1-l_1-\frac{l_2}{2}\right)\pi r_h^2}{\tau}.
\end{equation}
The radial component of the vector field is then
\begin{equation}
    \phi^{r_h}=\frac{\partial \mathcal{F}}{\partial r_h}=\left(1-l_1-\frac{l_2}{2}\right)\left(\frac{c_0}{2}-\frac{Q^2}{2 c_1 r_h^2}-\frac{\Lambda r_h^2}{2 c_2}\right)-\frac{2\left(1-l_1-\frac{l_2}{2}\right)\pi r_h}{\tau}.
\end{equation}
Solving $\phi^{r_h}=0$ gives the defect curve
\begin{equation}
    \tau(r_h) = \frac{4\pi r_h}{c_0 - \dfrac{Q^2}{c_1 r_h^2} - \dfrac{\Lambda r_h^2}{c_2}}.
\end{equation}

\begin{figure}[ht]
    \centering
    \includegraphics[width=1\textwidth]{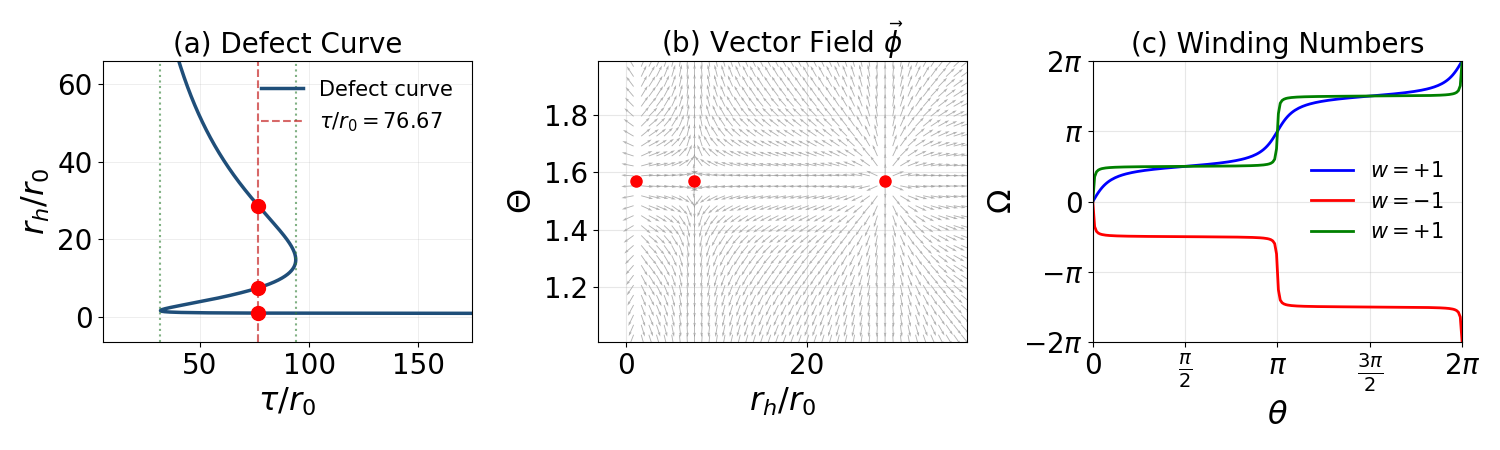}
    \caption{Defect Curves, zero distribution and winding numbers for Case B ($l_1=5\times10^{-10}$, $Q=3$, $\Lambda=-0.0005$ ).}
    \label{fig:a1}
\end{figure}

For Case B, with $l_2=-4l_1$, the defect curve is shown in Fig. \ref{fig:a1}a, and the vertical red line marks $\tau/r_0=76.67$. The curve contains one generation point and one annihilation point. For values of $\tau$ between these two points, three black hole branches coexist, corresponding to the van der Waals-type first-order transition in which the small and large black holes are connected through an unstable intermediate branch. Figure \ref{fig:a1}b shows the vector field $(\phi^{r_h},\phi^\Theta)$ in the $r_h$-$\Theta$ plane. The zeros occur at $r_h=3.30$, $22.69$, and $86.06$, with winding numbers $\omega=+1$, $-1$, and $+1$, respectively. The total topological number is $\sum_i\omega_i=+1$. Positive winding numbers identify the locally stable small and large black hole branches, whereas the negative winding number identifies the unstable intermediate branch.
\FloatBarrier
\FloatBarrier

\begin{figure}[ht]
    \centering
    \includegraphics[width=1\textwidth]{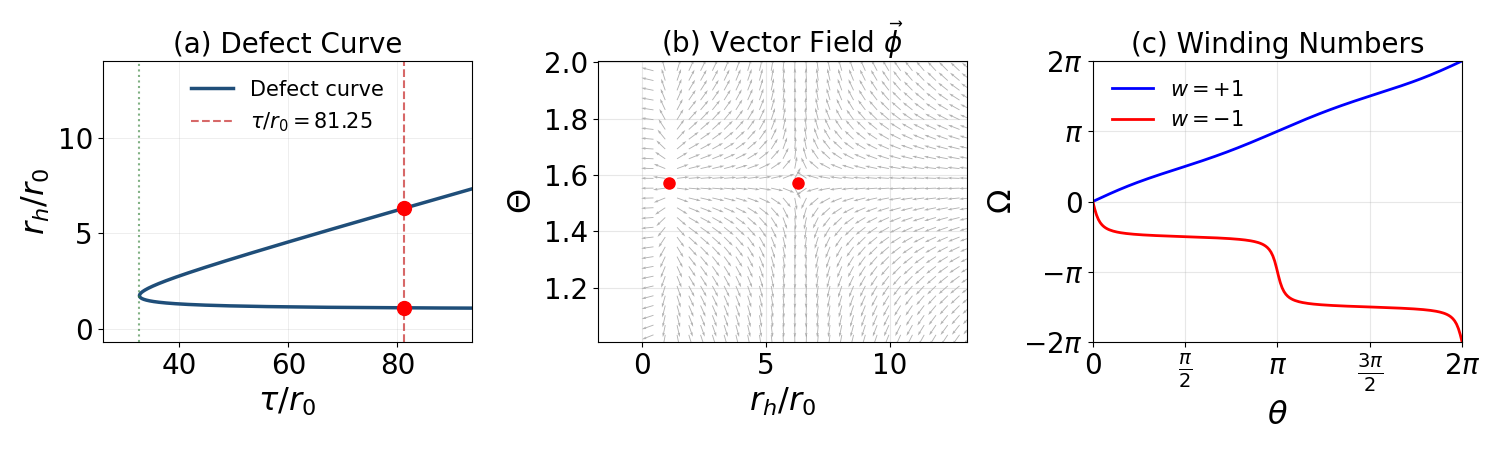}
    \caption{Defect Curves, zero distribution and winding numbers for Case A ($l_1=1\times10^{-10}$, $l_2=2\times10^{-10}$, $Q=0.8$).}
    \label{fig:a2}
\end{figure}

For Case A, where $\Lambda=0$, the defect curve in Fig. \ref{fig:a2}a has a single-fold structure with one generation point. For values of $\tau$  larger than the value at the generation point, two black hole branches coexist. The vertical dashed line in Fig. \ref{fig:a2}a marks $\tau/r_0=81.25$ at which the vector field and winding numbers are examined in Fig. \ref{fig:a2}b and \ref{fig:a2}c. At this value, Fig. \ref{fig:a2}b shows two zeros of $(\phi^{r_h},\phi^\Theta)$ at $r_h/r_0=0.88$ and $5.04$, with winding numbers $\omega=+1$ and $-1$, respectively. The total topological number is $\sum_i\omega_i=0$. Unlike Case B, there is no stable large black hole branch at large $r_h$, and the total topological number is zero.
\FloatBarrier
\FloatBarrier
Since $l_1$ and $l_2$ are independent in Case A, their individual effects can be studied directly. In both figures, the green solid, red dashed, and blue dash-dotted curves denote the relative deviation $\Delta r_h/r_{h,0}$ of the turning point radius, large black hole radius, and small black hole radius, respectively, at fixed cavity temperature $\tau$.

\begin{figure}[htbp]
    \centering
    \begin{subfigure}[b]{0.48\textwidth}
        \centering
        \includegraphics[width=\textwidth]{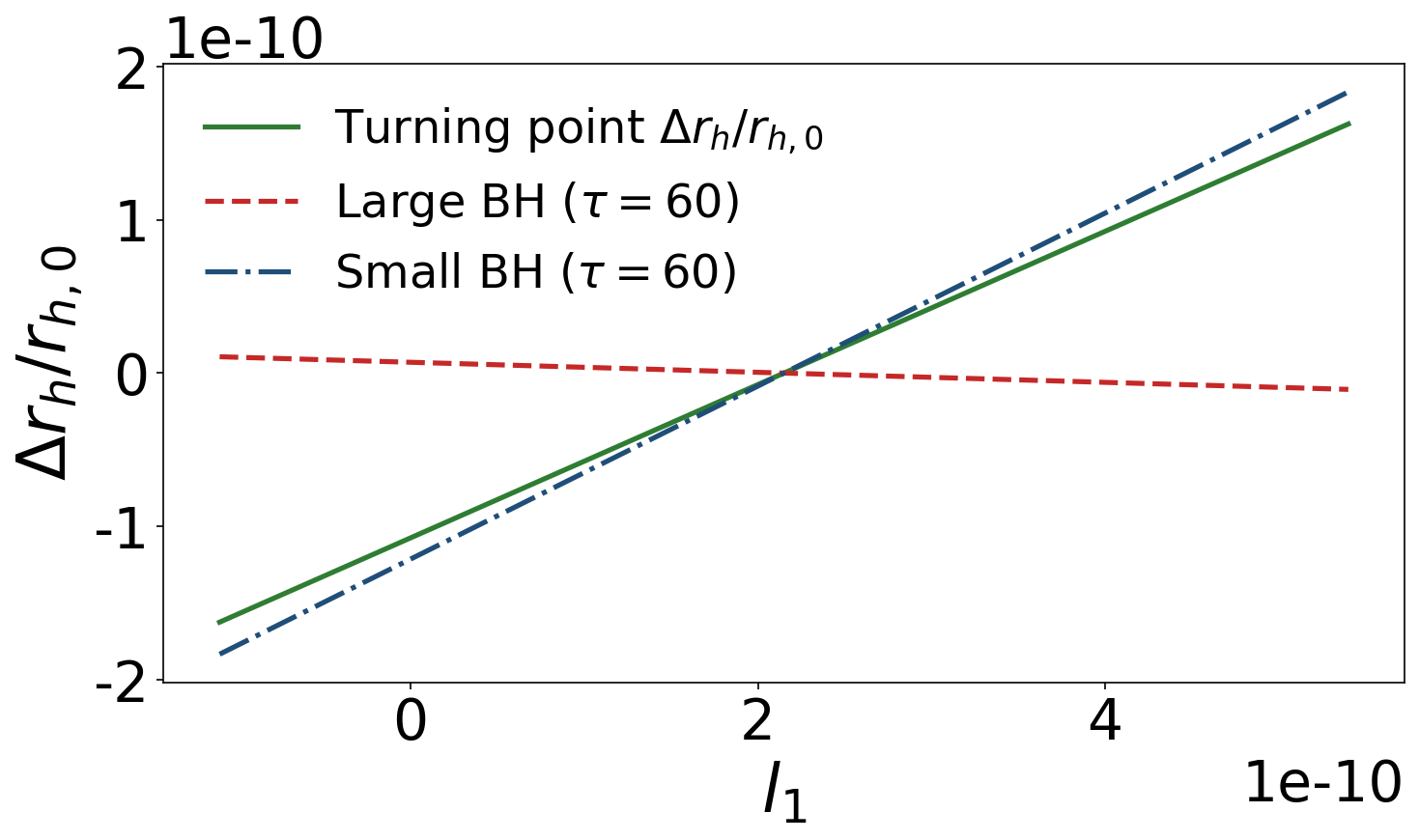}  
        \caption{}
        \label{fig:s1}
    \end{subfigure}
    \hfill  
    \begin{subfigure}[b]{0.48\textwidth}
        \centering
        \includegraphics[width=\textwidth]{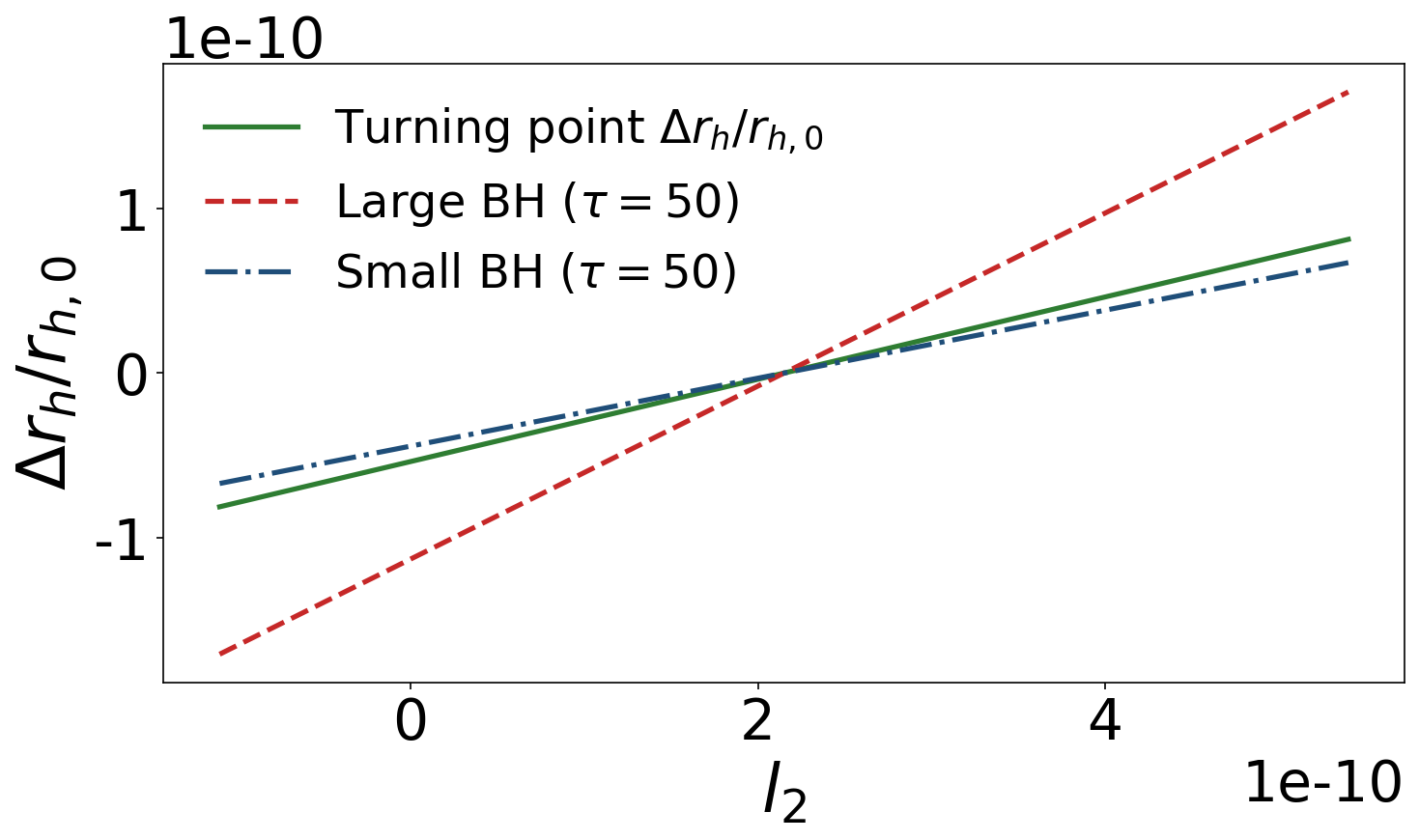}
        \caption{}
        \label{fig:s2}
    \end{subfigure}
    \vspace{0.3cm}
    
    \caption{Continuous evolution of the black hole horizon radius with the coupling parameters $l_1$  and $l_2$ in Case A. Panel \ref{fig:s1} fixes $l_2=2\times10^{-10}$ and varies $l_1$ over $[-1.1\times10^{-10},\,5.4\times10^{-10}]$. Panel \ref{fig:s2} fixes $l_1=1\times10^{-10}$ and varies $l_2$ over $[-1.1\times10^{-10},\,5.4\times10^{-10}]$. }
    \label{fig:main}
\end{figure}

In Fig. \ref{fig:s1}, as $l_1$ increases, the large black hole radius decreases slightly, while the small black hole radius and the turning point radius increase; the corresponding generation temperature $\tau_{\min}$ also increases. In Fig. \ref{fig:s2}, both horizon radii and the turning point radius increase with $l_2$, with the large black hole branch showing the strongest response, whereas $\tau_{\min}$ decreases.
\FloatBarrier
\FloatBarrier

\subsection{Critical Conditions from Defect Curves of Charged Kalb--Ramond Black Holes}

For the charged Kalb--Ramond AdS solution of Case B, the Joule--Thomson analysis and the defect-curve structure both indicate van der Waals-like thermodynamic behavior. We now rederive the critical temperature and pressure from the topological defect curve and compare them with the thermodynamic results of Sec. \ref{sec:thermodynamics}.

For Case B, $V(X)=\lambda X^2/2$ and $\Lambda\neq0$, with $l_2=-4l_1$. The defect curve relating the cavity parameter $\tau$ to the horizon radius is
\begin{equation}
\tau(r_h) = \frac{4\pi r_h}{c_0 - \dfrac{Q^2}{c_1 r_h^2} - \dfrac{\Lambda r_h^2}{c_2}},
\end{equation}
with
\begin{equation}
    c_0=\frac{1+l_1}{1+3l_1}, c_1=\frac{(1+l_1)^2}{1-l_1},  c_2=1+l_1.
\end{equation}
The turning points are determined by $d\tau/dr_h=0$, which gives
\begin{equation}
c_0 - \frac{3Q^2}{c_1 r_h^2} + \frac{\Lambda r_h^2}{c_2} = 0.
\label{eqX}
\end{equation}
Setting $x=r_h^2$ reduces this condition to
\begin{equation}
\frac{\Lambda}{c_2} x^2 + c_0 x - \frac{3Q^2}{c_1} = 0.
\label{eqX}
\end{equation}
The discriminant of Eq. \eqref{eqX} is
\begin{equation}
\Delta = c_0^2 + \frac{12\Lambda Q^2}{c_1 c_2}.
\end{equation}
Substituting the Case B coefficients gives
\begin{align}
c_0^2 &= \left(\frac{1+l_1}{1+3l_1}\right)^2 \\
\frac{1}{c_1 c_2} &= \frac{(1-l_1)}{(1+l_1)^2} \cdot \frac{1}{1+l_1} = \frac{(1-l_1)}{(1+l_1)^3},
\end{align}
and hence
\begin{equation}
\Delta = \frac{(1+l_1)^2}{(1+3l_1)^2} + \frac{12\Lambda (1-l_1)Q^2}{(1+l_1)^3} .
\end{equation}
The turning points exist when $\Delta\ge0$. The critical point is obtained from $\Delta=0$, yielding
\begin{equation}
\Lambda_c = -\frac{(1+l_1)^5}{12(1-l_1)(1+3l_1)^2 Q^2}.
\end{equation}
Using $P=-\Lambda/(8\pi)$, the critical pressure is
\begin{equation}
P_c = \frac{(1+l_1)^5}{96 \pi (1-l_1)(1+3l_1)^2 Q^2}.
\end{equation}
At $\Lambda=\Lambda_c$, Eq. \eqref{eqX} has a unique real root, corresponding to the critical radius
\begin{equation}
r_c = Q\sqrt{\frac{6(1-l_1)(1+3l_1)}{(1+l_1)^3}}.
\end{equation}
Substituting $r_c$ and $\Lambda_c$ into the defect curve gives the critical temperature
\begin{equation}
     T_{c}=\frac{1}{\tau(r_{c})}=\frac{(1+l_1)^{5/2}}{3\sqrt{6}\pi (1+3l_1)^{3/2}\sqrt{1-l_1}Q}.
\end{equation}
The critical quantities obtained from the defect curve agree with those derived from the van der Waals-like equation of state in Sec. \ref{sec:thermodynamics}. The pressure relative to $P_c$ does not change the total topological number，which is protected by the asymptotic behavior of the Hawking temperature at $r_h\to r_m$ and $r_h\to\infty$ \cite{Wei:2024gfz}, but it changes the local winding-number composition and therefore the phase-transition structure.

\begin{figure}[ht]
    \centering
    \includegraphics[width=0.5\textwidth]{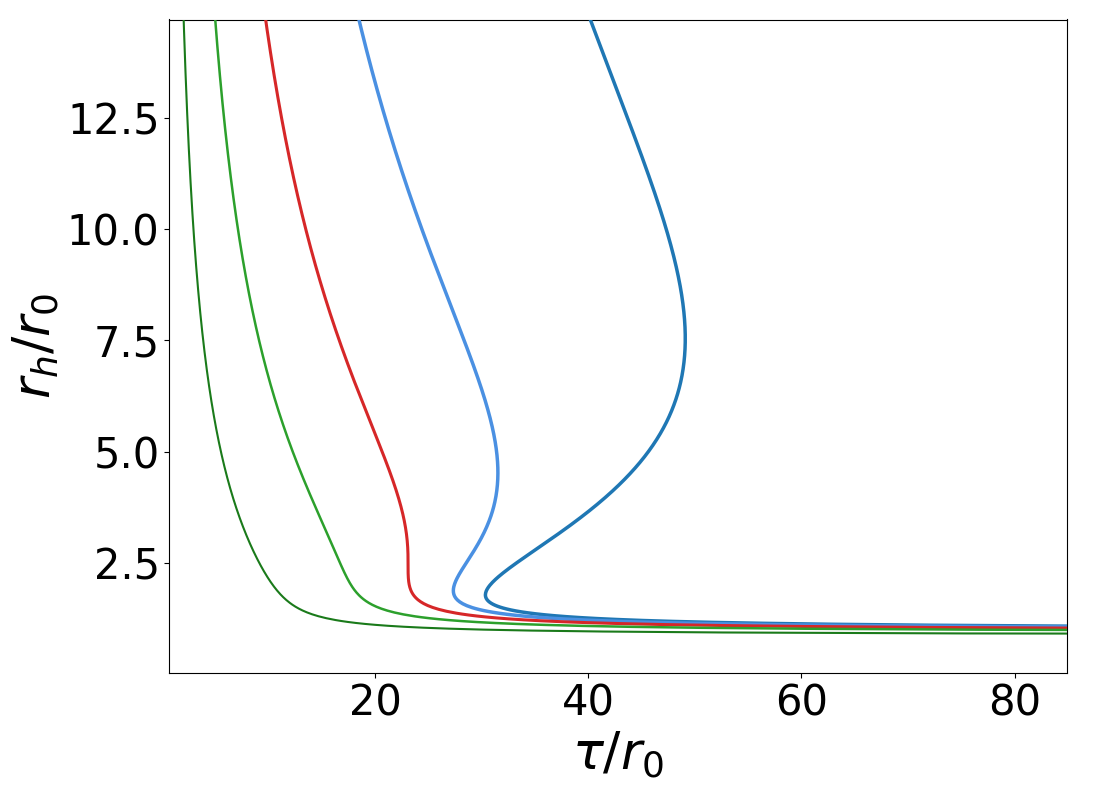}
    \vspace{0.3cm}
    \caption{Defect curves at various pressures for Case B with $l_1=1\times10^{-10}$, $Q=0.8$, and $P=1.03\times10^{-3},\,2.59\times10^{-3},\,5.18\times10^{-3},\,1.03\times10^{-2},\,2.59\times10^{-2}$.}
    \label{fig:22}
\end{figure}

For $l_1=1\times10^{-10}$ and $Q=0.8$, Fig. \ref{fig:22} illustrates the three regimes. For $\Lambda>\Lambda_c=-0.1302$ (equivalently $P<P_c=5.18\times10^{-3}$), the blue curves exhibit an S-shaped fold with one generation point and one annihilation point, corresponding to a first-order van der Waals-type transition. At $\Lambda=\Lambda_c$ ($P=P_c$), the two turning points merge into a single horizontal turning point, shown by the red curve. The latent heat then vanishes, $L=T\Delta S\to0$, and the transition becomes continuous. For $\Lambda<\Lambda_c$ ($P>P_c$), the green curves have no turning point and the defect curve is monotonic, indicating continuous state evolution rather than a phase transition.

\FloatBarrier
\FloatBarrier

\section{Conclusion}\label{sec:conclusion}
We have investigated static, spherically symmetric electrically charged black holes in a gravitational theory containing a background Kalb--Ramond field with a nonzero vacuum expectation value. We derived exact charged solutions, analyzed their thermodynamics using the Iyer--Wald formalism, studied the Joule--Thomson expansion, and examined the corresponding phase-transition structure within the topological framework of black hole thermodynamics.

First, following Ref. \cite{LWWL25}, we retained both nonminimal curvature couplings, including the scalar-curvature term $\xi_1B^{\mu\nu}B_{\mu\nu}R$, whose variation contributes nontrivially to the gravitational field equations. This led to charged Kalb--Ramond black hole solutions with and without a cosmological constant, extending and revising earlier results in which this coupling was not included \cite{DZY23}.

Second, the Iyer--Wald analysis showed that the Lorentz-violating couplings modify the energy and entropy. In the extended phase space, the charged Kalb--Ramond AdS solution displays van der Waals-like criticality and a well-defined Joule--Thomson expansion. Although the Kalb--Ramond parameter changes the individual values of $T_i^{\min}$ and $T_c$, it acts as a reparametrization of the thermodynamic phase space, so the dimensionless ratio $T_i^{\min}/T_c=1/2$ is preserved.

Third, the topological analysis distinguished the branch structure of the Case A and Case B solutions through their winding numbers. The charged Kalb--Ramond AdS solution has total topological number $+1$ and exhibits the standard small/intermediate/large black hole branch structure, whereas the $\Lambda=0$ solution has total topological number zero. The critical pressure and temperature were also recovered directly from the defect curve.

These results provide a basis for further studies of spontaneous Lorentz-symmetry-breaking effects in black hole physics, including possible observational probes through quasinormal modes, shadows, and gravitational-wave signatures.
\acknowledgments

We would like to thank Shao-Wen Wei for very useful discussions. This work was supported by the National Natural
Science Foundation of China (Grants No. 12475056 and No. 12247101), the Fundamental Research Funds for the Central Universities (Grants No. lzujbky-2025-it05 and lzujbky-2025-jdzx07), the Natural Science
Foundation of Gansu Province (No. 22JR5RA389, No.25JRRA799), Gansu Province's Top LeadingTalent Support
Plan, and the ‘111 Center’ under Grant No. B20063.

\bibliographystyle{JHEP}
\bibliography{ref}

@article{a1,
   title = {Spontaneous breaking of Lorentz symmetry in string theory},
  author = {Kosteleck\'y, V. Alan and Samuel, Stuart},
  journal = {Phys. Rev. D},
  volume = {39},
  issue = {2},
  pages = {683--685},
  numpages = {0},
  year = {1989},
  month = {Jan},
  publisher = {American Physical Society},
  doi = {10.1103/PhysRevD.39.683},
  url = {https://link.aps.org/doi/10.1103/PhysRevD.39.683},
}

@article{a2,
  title = {Loop quantum gravity and light propagation},
  author = {Alfaro, Jorge and Morales-T\'ecotl, Hugo A. and Urrutia, Luis F.},
  journal = {Phys. Rev. D},
  volume = {65},
  issue = {10},
  pages = {103509},
  numpages = {18},
  year = {2002},
  month = {Apr},
  publisher = {American Physical Society},
  doi = {10.1103/PhysRevD.65.103509},
  url = {https://link.aps.org/doi/10.1103/PhysRevD.65.103509},
}

@article{a3,
  doi = {10.1088/0264-9381/28/11/114012},
url = {https://doi.org/10.1088/0264-9381/28/11/114012},
year = {2011},
month = {may},
publisher = {},
volume = {28},
number = {11},
pages = {114012},
author = {Hořava, Petr},
title = {{General covariance in gravity at a Lifshitz point}},
journal = {Classical and Quantum Gravity},
}

@article{a4,
   title = {{Noncommutative Field Theory and Lorentz Violation}},
  author = {Carroll, Sean M. and Harvey, Jeffrey A. and Kosteleck\'y, V. Alan and Lane, Charles D. and Okamoto, Takemi},
  journal = {Phys. Rev. Lett.},
  volume = {87},
  issue = {14},
  pages = {141601},
  numpages = {4},
  year = {2001},
  month = {Sep},
  publisher = {American Physical Society},
  doi = {10.1103/PhysRevLett.87.141601},
  url = {https://link.aps.org/doi/10.1103/PhysRevLett.87.141601},
}

@article{a5,
  title = {{Gravity, Lorentz violation, and the standard model}},
  author = {Kosteleck\'y, V. Alan},
  journal = {Phys. Rev. D},
  volume = {69},
  issue = {10},
  pages = {105009},
  numpages = {20},
  year = {2004},
  month = {May},
  publisher = {American Physical Society},
  doi = {10.1103/PhysRevD.69.105009},
  url = {https://link.aps.org/doi/10.1103/PhysRevD.69.105009},
}

@article{a6,
  author       = {Lessa, L.A. and Silva, J.E.G. and Maluf, R.V. and Almeida, C.A.S.},
  journal      = {Eur. Phys. J. C},
  title        = {{Modified black hole solution with a background Kalb-Ramond field}},
  year         = {2020},
  volume       = {80},
  pages        = {335},
  eprint       = {1911.10296},
  archiveprefix = {arXiv},
  primaryclass = {gr-qc},
  url = {https://doi.org/10.1140/epjc/s10052-020-7902-1},
}

@article{a7,
  author       = {Atamurotov, F. and Ortiqboev, D. and Abdujabbarov, A. and Mustafa, G.},
  journal      = {Eur. Phys. J. C},
  title        = {{Particle dynamics and gravitational weak lensing around black hole in the Kalb-Ramond gravity}},
  year         = {2022},
  volume       = {82},
  pages        = {659},
  url = {https://doi.org/10.1140/epjc/s10052-022-10619-z},
}

@article{a8,
  title = {{Traversable wormhole solution with a background Kalb–Ramond field}},
journal = {Annals of Physics},
volume = {433},
pages = {168604},
year = {2021},
issn = {0003-4916},
doi = {10.1016/j.aop.2021.168604},
url = {https://www.sciencedirect.com/science/article/pii/S0003491621002104},
author = {L.A. Lessa and R. Oliveira and J.E.G. Silva and C.A.S. Almeida},
}

@article{a9,
  author       = {Maluf, R.V. and Muniz, C.R.},
  journal      = {Eur. Phys. J. C},
  title        = {{Exact solution for a traversable wormhole in a curvature-coupled antisymmetric background field}},
  year         = {2022},
  volume       = {82},
  pages        = {445},
  eprint       = {2110.12202},
  archiveprefix = {arXiv},
  primaryclass = {gr-qc},
  url = {https://doi.org/10.1140/epjc/s10052-022-10409-7},
}

@article{a10,
doi = {10.1088/1475-7516/2024/05/029},
url = {https://doi.org/10.1088/1475-7516/2024/05/029},
year = {2024},
month = {may},
publisher = {IOP Publishing},
volume = {2024},
number = {05},
pages = {029},
author = {Araújo Filho, A.A. and Reis, J.A.A.S. and Hassanabadi, H.},
title = {Exploring antisymmetric tensor effects on black hole shadows and quasinormal frequencies},
journal = {Journal of Cosmology and Astroparticle Physics},
}

@article{ABK10,
  title = {Lorentz violation with an antisymmetric tensor},
  author = {Altschul, Brett and Bailey, Quentin G. and Kosteleck\'y, V. Alan},
  journal = {Phys. Rev. D},
  volume = {81},
  issue = {6},
  pages = {065028},
  numpages = {19},
  year = {2010},
  month = {Mar},
  publisher = {American Physical Society},
  doi = {10.1103/PhysRevD.81.065028},
  url = {https://link.aps.org/doi/10.1103/PhysRevD.81.065028},
}

@article{HY01,
  title = {{Spontaneous Lorentz symmetry breaking by an antisymmetric tensor field}},
  author = {Higashijima, Kiyoshi and Yokoi, Naoto},
  journal = {Phys. Rev. D},
  volume = {64},
  issue = {2},
  pages = {025004},
  numpages = {8},
  year = {2001},
  month = {Jun},
  publisher = {American Physical Society},
  doi = {10.1103/PhysRevD.64.025004},
  url = {https://link.aps.org/doi/10.1103/PhysRevD.64.025004},
}

@article{MCAA19,
  doi = {10.1209/0295-5075/124/61001},
url = {https://doi.org/10.1209/0295-5075/124/61001},
year = {2019},
month = {jan},
publisher = {EDP Sciences, IOP Publishing and Società Italiana di Fisica},
volume = {124},
number = {6},
pages = {61001},
author = {Maluf, R. V. and Araújo Filho, A. A. and Cruz, W. T. and Almeida, C. A. S.},
title = {{Antisymmetric tensor propagator with spontaneous Lorentz violation}},
journal = {Europhysics Letters},
}

@article{MS99,
  doi = {10.1088/0264-9381/16/12/102},
url = {https://doi.org/10.1088/0264-9381/16/12/102},
year = {1999},
month = {dec},
publisher = {},
volume = {16},
number = {12},
pages = {L89},
author = {Parthasarathi Majumdar and Soumitra SenGupta},
title = {Parity-violating gravitational coupling of electromagnetic 
fields},
journal = {Classical and Quantum Gravity},
}

@article{DZY23,
 author       = {Duan, Z. Q. and Zhao, J. Y. and Yang, K.},
  journal      = {Eur. Phys. J. C},
  title        = {{Electrically charged black holes in gravity with a background Kalb--Ramond field}},
  year         = {2024},
  volume       = {84},
  pages        = {798},
  doi          = {10.1140/epjc/s10052-024-13188-5},
  url = {https://doi.org/10.1140/epjc/s10052-024-13188-5},
}

@article{LSMA20,
  author       = {Lessa, L.A. and Silva, J.E.G. and Maluf, R.V. and Almeida, C.A.S.},
  journal      = {Eur. Phys. J. C},
  title        = {{Modified black hole solution with a background Kalb-Ramond field}},
  year         = {2020},
  volume       = {80},
  pages        = {335},
  eprint       = {1911.10296},
  archiveprefix = {arXiv},
  primaryclass = {gr-qc},
  url = {https://doi.org/10.1140/epjc/s10052-020-7902-1},
}

@article{LWWL25,
  doi = {10.1088/1475-7516/2025/11/056},
url = {https://doi.org/10.1088/1475-7516/2025/11/056},
year = {2025},
month = {nov},
publisher = {IOP Publishing},
volume = {2025},
number = {11},
pages = {056},
author = {Liu, Jia-Zhou and Wu, Shan-Ping and Wei, Shao-Wen and Liu, Yu-Xiao},
title = {{Exact black hole solutions in gravity with a background Kalb-Ramond field}},
journal = {Journal of Cosmology and Astroparticle Physics},
}

@article{PhysRevD.108.124004,
  title = {{Static and spherically symmetric black holes in gravity with a background Kalb-Ramond field}},
  author = {Yang, Ke and Chen, Yue-Zhe and Duan, Zheng-Qiao and Zhao, Ju-Ying},
  journal = {Phys. Rev. D},
  volume = {108},
  issue = {12},
  pages = {124004},
  numpages = {11},
  year = {2023},
  month = {Dec},
  publisher = {American Physical Society},
  doi = {10.1103/PhysRevD.108.124004},
  url = {https://link.aps.org/doi/10.1103/PhysRevD.108.124004},
}

@article{PhysRevD.40.1886,
  title = {Gravitational phenomenology in higher-dimensional theories and strings},
  author = {Kosteleck\'y, V. Alan and Samuel, Stuart},
  journal = {Phys. Rev. D},
  volume = {40},
  issue = {6},
  pages = {1886--1903},
  numpages = {0},
  year = {1989},
  month = {Sep},
  publisher = {American Physical Society},
  doi = {10.1103/PhysRevD.40.1886},
  url = {https://link.aps.org/doi/10.1103/PhysRevD.40.1886},
}

@article{r1,
  author       = {Xiao, Yong and Liu, Yu-Xiao and Tian, Yu and Zhang, Hongbao},
  title        = {{Explicit and covariant formula for thermodynamic volume in extended black hole thermodynamics}},
  year         = {2025},
  eprint       = {2512.01916},
  archiveprefix = {arXiv},
  primaryclass = {gr-qc},
  journal = {},
}

@article{r2,
   title = {{Extended Black Hole Thermodynamics from Extended Iyer-Wald Formalism}},
  author = {Xiao, Yong and Tian, Yu and Liu, Yu-Xiao},
  journal = {Phys. Rev. Lett.},
  volume = {132},
  issue = {2},
  pages = {021401},
  numpages = {7},
  year = {2024},
  month = {Jan},
  publisher = {American Physical Society},
  doi = {10.1103/PhysRevLett.132.021401},
  url = {https://link.aps.org/doi/10.1103/PhysRevLett.132.021401},
}

@article{t11,
  title = {{Nonexistence of Baryon Number for Static Black Holes}},
  author = {Bekenstein, Jacob D.},
  journal = {Phys. Rev. D},
  volume = {5},
  issue = {6},
  pages = {1239--1246},
  numpages = {0},
  year = {1972},
  month = {Mar},
  publisher = {American Physical Society},
  doi = {10.1103/PhysRevD.5.1239},
  url = {https://link.aps.org/doi/10.1103/PhysRevD.5.1239},
}

@article{t12,
  author       = {Bekenstein, J. D.},
  journal      = {Phys. Rev. D},
  title        = {{Black holes and entropy}},
  year         = {1973},
  volume       = {7},
  pages        = {2333},
}

@article{t14,
  author       = {Bekenstein, J. D.},
  journal      = {Phys. Rev. D},
  title        = {{Generalized second law of thermodynamics}},
  year         = {1974},
  volume       = {9},
  pages        = {3292},
}

@article{t15,
  author       = {Hawking, S. W.},
  journal      = {Nature},
  title        = {{Black Hole Explosions?}},
  year         = {1974},
  volume       = {248},
  pages        = {30},
}

@article{t16,
  author       = {Hawking, S. W.},
  journal      = {Commun. Math. Phys.},
  title        = {{Particle Creation by Black Holes}},
  year         = {1975},
  volume       = {43},
  pages        = {199},
}

@article{OA17,
  author       = {\"{O}kc\"{u}, \"{O}. and Ayd{\i}ner, E.},
  journal      = {Eur. Phys. J. C},
  title        = {{Joule--Thomson expansion of the charged AdS black holes}},
  year         = {2017},
  volume       = {77},
  pages        = {24},
  doi          = {10.1140/epjc/s10052-017-4598-y},
  url = {https://doi.org/10.1140/epjc/s10052-017-4598-y},
}

@article{Hegde:2024jtgaussbonnet,
    author = {Hegde, Kartheek and Ahmed Rizwan, C. L. and Ajith, K. M. and Naveena Kumara, A. and Ali, Md Sabir and Punacha, Shreyas},
    title = {{Thermodynamics, phase transition and Joule--Thomson expansion of 4-D Gauss--Bonnet AdS black hole}},
    doi = {10.1142/S0217751X24500805},
    journal = {International Journal of Modern Physics A},
    volume = {39},
    number = {21},
    pages = {2450080},
    year = {2024}
}

@article{Kruglov:2022nedadsjt,
    author = {Kruglov, S. I.},
    title = {{NED--AdS black holes, extended phase space thermodynamics and Joule--Thomson expansion}},
    doi = {10.1016/j.nuclphysb.2022.115949},
    journal = {Nuclear Physics B},
    volume = {984},
    pages = {115949},
    year = {2022}
}

@article{ElHadri:2026noncommutativernads,
    author = "El Hadri, W. and Jemri, M.",
    title = {{Thermodynamics and Criticality of Noncommutative RN--AdS Black Holes}},
    doi = {10.1007/s13538-025-01940-5},
    journal = "Brazilian Journal of Physics",
    volume = "56",
    pages = "23",
    year = "2026",
}

@article{Kruglov:2022nonlinearchargedjt,
    author = "Kruglov, S. I.",
    title = {{Nonlinearly charged AdS black holes, extended phase space thermodynamics and Joule--Thomson expansion}},
    doi = {10.1016/j.aop.2022.168894},
    journal = "Annals of Physics",
    volume = "441",
    pages = "168894",
    year = "2022",
}

@article{Media:2025kerrdslorentzviolationjt,
    author = "Media, N. and Singh, T. I.",
    title = {{Joule--Thomson Expansion of Kerr-Newman-de Sitter Black Hole Under Lorentz Violation Theory}},
    doi = {10.1007/s10773-025-05949-z},
    journal = "International Journal of Theoretical Physics",
    volume = "64",
    pages = "82",
    year = "2025",
}

@article{Ahmed:2025oxp,
  author       = {Ahmed, Faizuddin and Silva, Edilberto O.},
  journal      = {Chinese J. Phys.},
  title        = {{Thermodynamics and P-V criticality of charged AdS black holes with a cloud of strings in Kalb-Ramond gravity}},
  year         = {2026},
  doi          = {10.1016/j.cjph.2026.06.007},
}

@article{Kruglov:2022symmetrynedthermo,
    author = "Kruglov, Sergey Il’ich",
    title = {{AdS Black Holes in the Framework of Nonlinear Electrodynamics, Thermodynamics, and Joule--Thomson Expansion}},
    doi = {10.3390/sym14081597},
    journal = "Symmetry",
    volume = "14",
    number = "8",
    pages = "1597",
    year = "2022",
}

@article{Biswas:2021yangmillsjt,
    author = "Biswas, Anindya",
    title = {{Joule-Thomson expansion of AdS black holes in Einstein Power-Yang-mills gravity}},
    doi = {10.1088/1402-4896/ac2b42},
    journal = "Physica Scripta",
    volume = "96",
    number = "12",
    pages = "125310",
    year = "2021",
}

@article{Ali:2025nedrnadsjt,
    author = "Ali, R. H. and Abbas, G. and Jawad, Abdul and Alkahtani, Badr S. and Mustafa, G.",
    title = {{Mathematical formalism of Joule--Thomson process for ADS--RN black hole coupled with non-linear electrodynamics field}},
    doi = {10.1016/j.nuclphysb.2024.116735},
    journal = "Nuclear Physics B",
    volume = "1010",
    pages = "116735",
    year = "2025",
}

@article{Kruglov:2023gravcosmolmagneticjt,
    author = "Kruglov, S. I.",
    title = {{Magnetically Charged AdS Black Holes and Joule--Thomson Expansion}},
    doi = {10.1134/S0202289323010073},
    journal = "Gravitation and Cosmology",
    volume = "29",
    pages = "57–61",
    year = "2023",
}

@article{Kruglov:2023canadianjpmagnetic,
    author = "Kruglov, S. I.",
    title = {{Magnetic black holes within Einstein--AdS gravity coupled to nonlinear electrodynamics, extended phase space thermodynamics and Joule--Thomson expansion}},
    doi = {10.1139/cjp-2023-0119},
    journal = "Canadian Journal of Physics",
    volume = "101",
    number = "12",
    pages = "739–748",
    year = "2023",
}

@article{Wang:2025quantumcorrectedadsjt,
    author = "Wang, Rui-Bo and You, Lei and Ma, Shi-Jie and Deng, Jian-Bo and Hu, Xian-Ru",
    title = {{Thermodynamic phase transition and Joule--Thomson expansion of a quantum corrected black hole in AdS spacetime}},
    doi = {10.1088/1674-1137/ade4a9},
    journal = "Chinese Physics C",
    volume = "49",
    number = "11",
    pages = "115102",
    year = "2025",
}

@article{Cao:2021darkmatterrnadsjt,
    author = "Cao, Yihe and Feng, Hanwen and Hong, Wei and Tao, Jun",
    title = "{Joule--Thomson expansion of RN--AdS black hole immersed in perfect fluid dark matter}",
    doi = {10.1088/1572-9494/ac1066},
    journal = "Communications in Theoretical Physics",
    volume = "73",
    number = "9",
    pages = "095403",
    year = "2021",
}

@article{Liang:2021torusblackholejt,
    author = "Liang, J. and Lin, W. and Mu, B.",
    title = "{Joule--Thomson expansion of the torus--like black hole}",
    doi = {10.1140/epjp/s13360-021-02119-y},
    journal = "European Physical Journal Plus",
    volume = "136",
    pages = "1169",
    year = "2021",
}

@article{Sekhmani:20225drchargedjt,
    author = "Sekhmani, Yassine and Dahbi, Zakaria and Najim, Abdelhafid and Waqdim, Abderrahman",
    title = "{Joule--Thomson expansion of 5-dimensional R-charged black holes}",
    doi = {10.1016/j.aop.2022.169060},
    journal = "Annals of Physics",
    volume = "444",
    pages = "169060",
    year = "2022",
}

@article{Masmar:2023nonlinearchargedjt,
    author = "Masmar, K.",
    title = "{Joule--Thomson expansion for a nonlinearly charged Anti-de Sitter black hole}",
    doi = {10.1142/S0219887823500809},
    journal = "International Journal of Geometric Methods in Modern Physics",
    volume = "20",
    number = "05",
    pages = "2350080",
    year = "2023",
}

@article{Mo:2018ddimchargedjt,
    author = "Mo, Jie-Xiong and Li, Gu-Qiang and Lan, Shan-Quan and Xu, Xiao-Bao",
    title = "{Joule--Thomson expansion of d-dimensional charged AdS black holes}",
    doi = {10.1103/PhysRevD.98.124032},
    journal = "Phys. Rev. D",
    volume = "98",
    number = "12",
    pages = "124032",
    year = "2018",
}

@article{Alipour:2025yangmillskerrsenjt,
    author = "Alipour, M. R. and Gashti, S. N. and Afshar, M. A. S. and others",
    title = "{Cooling and heating regions of Joule--Thomson expansion for AdS black holes: Einstein-Maxwell-power-Yang-Mills and Kerr Sen black holes}",
    doi = {10.1007/s10714-025-03393-2},
    journal = "General Relativity and Gravitation",
    volume = "57",
    pages = "61",
    year = "2025",
}

@article{a12,
  doi = {10.1088/0264-9381/6/12/018},
url = {https://doi.org/10.1088/0264-9381/6/12/018},
year = {1989},
month = {dec},
publisher = {},
volume = {6},
number = {12},
pages = {1909},
author = {P C W Davies},
title = {{Thermodynamic phase transitions of Kerr-Newman black holes in de Sitter space}},
journal = {Classical and Quantum Gravity},
}

@article{a13,
  author       = {Hawking, S. W. and Page, D. N.},
  journal      = {Commun. Math. Phys.},
  title        = {{Thermodynamics of Black Holes in anti-De Sitter Space}},
  year         = {1983},
  volume       = {87},
  pages        = {577},
  doi          = {10.1007/BF01208266},
  url = {https://doi.org/10.1007/BF01208266},
}

@article{a14,
  author       = {Curir, A.},
  journal      = {Gen. Rel. Grav.},
  title        = {{Rotating black holes as dissipative spin-thermodynamical systems}},
  year         = {1981},
  volume       = {13},
  pages        = {417},
  doi          = {10.1007/BF00756588},
   url = {https://doi.org/10.1007/BF00756588},
}

@article{a15,
  author       = {Curir, Anna},
  journal      = {Gen. Rel. Grav.},
  title        = {{Black hole emissions and phase transitions}},
  year         = {1981},
  volume       = {13},
  pages        = {1177},
  doi          = {10.1007/BF00759866},
  url = {https://doi.org/10.1007/BF00759866},
}

@article{a16,
  title = {Nonequilibrium thermodynamic fluctuations of black holes},
  author = {Pav\'on, D. and Rub\'{\i}, J. M.},
  journal = {Phys. Rev. D},
  volume = {37},
  issue = {8},
  pages = {2052--2058},
  numpages = {0},
  year = {1988},
  month = {Apr},
  publisher = {American Physical Society},
  doi = {10.1103/PhysRevD.37.2052},
  url = {https://link.aps.org/doi/10.1103/PhysRevD.37.2052},
}

@article{a17,
   title = {{Phase transition in Reissner-Nordstr\"om black holes}},
  author = {Pav\'on, Diego},
  journal = {Phys. Rev. D},
  volume = {43},
  issue = {8},
  pages = {2495--2497},
  numpages = {0},
  year = {1991},
  month = {Apr},
  publisher = {American Physical Society},
  doi = {10.1103/PhysRevD.43.2495},
  url = {https://link.aps.org/doi/10.1103/PhysRevD.43.2495},
}

@article{a18,
  author       = {Kaburaki, O.},
  journal      = {Gen. Rel. Grav.},
  title        = {{Critical behavior of extremal Kerr-Newman black holes}},
  year         = {1996},
  volume       = {28},
  pages        = {843},
  url = {https://doi.org/10.1007/BF02104753},
}

@article{a19,
  title = {Critical behavior in (2+1)-dimensional black holes},
  author = {Cai, Rong-Gen and Lu, Zhi-Jiang and Zhang, Yuan-Zhong},
  journal = {Phys. Rev. D},
  volume = {55},
  issue = {2},
  pages = {853--860},
  numpages = {0},
  year = {1997},
  month = {Jan},
  publisher = {American Physical Society},
  doi = {10.1103/PhysRevD.55.853},
  url = {https://link.aps.org/doi/10.1103/PhysRevD.55.853},
}

@article{a20,
  title = {{Thermodynamic curvature of the BTZ black hole}},
  author = {Cai, Rong-Gen and Cho, Jin-Ho},
  journal = {Phys. Rev. D},
  volume = {60},
  issue = {6},
  pages = {067502},
  numpages = {4},
  year = {1999},
  month = {Aug},
  publisher = {American Physical Society},
  doi = {10.1103/PhysRevD.60.067502},
  url = {https://link.aps.org/doi/10.1103/PhysRevD.60.067502},
}

@article{a21,
  title = {{Thermodynamic critical and geometrical properties of charged BTZ black hole}},
  author = {Wei, Yi-Huan},
  journal = {Phys. Rev. D},
  volume = {80},
  issue = {2},
  pages = {024029},
  numpages = {8},
  year = {2009},
  month = {Jul},
  publisher = {American Physical Society},
  doi = {10.1103/PhysRevD.80.024029},
  url = {https://link.aps.org/doi/10.1103/PhysRevD.80.024029},
}

@article{a22,
  title = {General framework to study the extremal phase transition of black holes},
  author = {Bhattacharya, Krishnakanta and Dey, Sumit and Majhi, Bibhas Ranjan and Samanta, Saurav},
  journal = {Phys. Rev. D},
  volume = {99},
  issue = {12},
  pages = {124047},
  numpages = {15},
  year = {2019},
  month = {Jun},
  publisher = {American Physical Society},
  doi = {10.1103/PhysRevD.99.124047},
  url = {https://link.aps.org/doi/10.1103/PhysRevD.99.124047},
}

@article{a23,
  doi = {10.1088/0264-9381/26/19/195011},
url = {https://doi.org/10.1088/0264-9381/26/19/195011},
year = {2009},
month = {sep},
publisher = {},
volume = {26},
number = {19},
pages = {195011},
author = {Kastor, David and Ray, Sourya and Traschen, Jennie},
title = {{Enthalpy and the mechanics of AdS black holes}},
journal = {Classical and Quantum Gravity},
}

@article{a24,
  author       = {Dolan, B. P.},
  journal      = {Class. Quant. Grav.},
  title        = {{The cosmological constant and the black hole equation of state}},
  year         = {2011},
  volume       = {28},
  pages        = {125020},
  doi          = {10.1088/0264-9381/28/12/125020},
  eprint       = {1008.5023},
  archiveprefix = {arXiv},
  primaryclass = {gr-qc},
}

@article{a25,
  doi = {10.1088/0264-9381/28/23/235017},
url = {https://doi.org/10.1088/0264-9381/28/23/235017},
year = {2011},
month = {nov},
publisher = {IOP Publishing},
volume = {28},
number = {23},
pages = {235017},
author = {Dolan, Brian P},
title = {Pressure and volume in the first law of black hole thermodynamics},
journal = {Classical and Quantum Gravity},
}

@article{a26,
  title = {Compressibility of rotating black holes},
  author = {Dolan, Brian P.},
  journal = {Phys. Rev. D},
  volume = {84},
  issue = {12},
  pages = {127503},
  numpages = {3},
  year = {2011},
  month = {Dec},
  publisher = {American Physical Society},
  doi = {10.1103/PhysRevD.84.127503},
  url = {https://link.aps.org/doi/10.1103/PhysRevD.84.127503},
}

@incollection{a27,
  author = {Brian P. Dolan},
title = {{Where is the PdV in the First Law of Black Hole Thermodynamics?}},
booktitle = {Open Questions in Cosmology},
publisher = {IntechOpen},
address = {London},
year = {2012},
editor = {Gonzalo J. Olmo},
chapter = {12},
doi = {10.5772/52455},
url = {https://doi.org/10.5772/52455},
}

@article{a28,
  author       = {Kubiznak, D. and Mann, R. B.},
  journal      = {JHEP},
  title        = {{P-V criticality of charged AdS black holes}},
  year         = {2012},
  volume       = {07},
  pages        = {033},
  doi          = {10.1007/JHEP07(2012)033},
  eprint       = {1205.0559},
  archiveprefix = {arXiv},
  primaryclass = {hep-th},
  url = {https://doi.org/10.1007/JHEP07(2012)033},
}

@article{a29,
  doi = {10.1088/1361-6382/aa5c69},
url = {https://doi.org/10.1088/1361-6382/aa5c69},
year = {2017},
month = {feb},
publisher = {IOP Publishing},
volume = {34},
number = {6},
pages = {063001},
author = {Kubizňák, David and Mann, Robert B and Teo, Mae},
title = {{Black hole chemistry: thermodynamics with Lambda}},
journal = {Classical and Quantum Gravity},
}

@article{a30,
  title = {{van der Waals criticality in AdS black holes: A phenomenological study}},
  author = {Bhattacharya, Krishnakanta and Majhi, Bibhas Ranjan and Samanta, Saurav},
  journal = {Phys. Rev. D},
  volume = {96},
  issue = {8},
  pages = {084037},
  numpages = {8},
  year = {2017},
  month = {Oct},
  publisher = {American Physical Society},
  doi = {10.1103/PhysRevD.96.084037},
  url = {https://link.aps.org/doi/10.1103/PhysRevD.96.084037},
}

@article{Wei:2022dzw,
    author = "Wei, Shao-Wen and Liu, Yu-Xiao and Mann, Robert B.",
    title = "{Black Hole Solutions as Topological Thermodynamic Defects}",
    eprint = "2208.01932",
    archivePrefix = "arXiv",
    primaryClass = "gr-qc",
    doi = {10.1103/PhysRevLett.129.191101},
    journal = "Phys. Rev. Lett.",
    volume = "129",
    number = "19",
    pages = "191101",
    year = "2022",
}

@article{Wei:2024gfz,
    author = "Wei, Shao-Wen and Liu, Yu-Xiao and Mann, Robert B.",
    title = "{Universal topological classifications of black hole thermodynamics}",
    eprint = "2409.09333",
    archivePrefix = "arXiv",
    primaryClass = "gr-qc",
    doi = {10.1103/PhysRevD.110.L081501},
    journal = "Phys. Rev. D",
    volume = "110",
    number = "8",
    pages = "L081501",
    year = "2024",
}

@article{Wei2026,
  author       = {Wei, Shao-Wen and Liu, Yu-Xiao},
  journal      = {Sci. China Phys. Mech. Astron.},
  title        = {{Topology of black hole thermodynamics: A brief review}},
  year         = {2026},
  volume       = {69},
  number       = {6},
  pages        = {260401},
  eprint       = {2605.00037},
  archiveprefix = {arXiv},
  primaryclass = {gr-qc},
  url={https://doi.org/10.1007/s11433-025-2923-3},
}

@article{York,
  title = {{Black-hole thermodynamics and the Euclidean Einstein action}},
  author = {York, James W.},
  journal = {Phys. Rev. D},
  volume = {33},
  issue = {8},
  pages = {2092--2099},
  numpages = {0},
  year = {1986},
  month = {Apr},
  publisher = {American Physical Society},
  doi = {10.1103/PhysRevD.33.2092},
  url = {https://link.aps.org/doi/10.1103/PhysRevD.33.2092},
}

@incollection{Duan1,
  author = {Yi-Shi Duan and Mo-Lin Ge},
  title = {{SU(2) Gauge Theory and Electrodynamics with N Magnetic Monopoles}},
  booktitle = {Memorial Volume for Yi-Shi Duan},
  chapter = {Chapter 1},
  pages = {1-15},
  doi = {10.1142/9789813237278_0001},
  publisher = {World Scientific},
  year = {2018},
}

@techreport{Duan2,
  author      = {Duan, Y.-S.},
  title       = {{The structure of the topological current}},
  year        = {1984},
  institution = {SLAC},
  number      = {SLAC-PUB-3301},
}

@article{by1,
  author       = {Liu, J.Z. and Wu, S.P. and Wei, S.W. and others},
  journal      = {Sci. China Phys. Mech. Astron.},
  title        = {{Exact black hole solutions in bumblebee gravity with lightlike or spacelike VEVs}},
  year         = {2026},
  volume       = {69},
  pages        = {270411},
  doi          = {10.1007/s11433-026-2961-8},
}

@article{by2,
	author={Gu, Yun-Tao and Guo, Wen-Di and LIU, Yu-Xiao},
	title={{Quasinormal modes of an electrically charged Kalb--Ramond black hole}},
	journal={Chinese Physics C},
	url={http://iopscience.iop.org/article/10.1088/1674-1137/ae6ed3},
	year={2026},
}

@article{by3,
  title = {{Stability analysis of cosmological perturbations in the bumblebee model: Parameter constraints and gravitational waves}},
  author = {Lai, Xiao-Bin and Dong, Yu-Qi and Fan, Yu-Zhi and Liu, Yu-Xiao},
  journal = {Phys. Rev. D},
  volume = {113},
  issue = {4},
  pages = {044003},
  numpages = {23},
  year = {2026},
  month = {Feb},
  publisher = {American Physical Society},
  doi = {10.1103/q6fk-3lkj},
  url = {https://link.aps.org/doi/10.1103/q6fk-3lkj},
}

@article{by4,
    author = "Li, Bo-Rui and Liu, Jia-Zhou and Guo, Wen-Di and Liu, Yu-Xiao",
    title = {{Quasinormal modes of a charged spherically symmetric black hole in bumblebee gravity}},
    eprint = "2510.20503",
    archivePrefix = "arXiv",
    primaryClass = "gr-qc",
    month = "10",
    year = "2025",
    journal={},    
}

@article{by5,
title = {{Quasinormal modes of spherically symmetric black hole with cosmological constant and global monopole in bumblebee gravity}},
journal = {Nuclear Physics B},
volume = {1018},
pages = {117006},
year = {2025},
issn = {0550-3213},
doi = {https://doi.org/10.1016/j.nuclphysb.2025.117006},
url = {https://www.sciencedirect.com/science/article/pii/S0550321325002159},
author = {Yenshembam {Priyobarta Singh} and Irengbam {Roshila Devi} and Telem {Ibungochouba Singh}},
}

@article{by6,
    author = "Lin, Yu-Xuan and Liu, Jia-Zhou and Liu, Yu-Xiao",
    title = {{Dyonic Black Holes in Lorentz--Violating Gravity with a Background Kalb--Ramond Field}},
    eprint = "2605.18371",
    archivePrefix = "arXiv",
    primaryClass = "gr-qc",
    month = "5",
    year = "2026",
    journal={},  
}

@article{by7,
    author = "Liu, Hui-Fa and Liu, Wentao and Liu, Yu-Xiao and Su, Qi and Zeng, Ding-fang",
    title = {{Gravitational--Bumblebee perturbations: Exact decoupling and isospectrality}},
    eprint = "2605.02820",
    archivePrefix = "arXiv",
    primaryClass = "gr-qc",
    month = "5",
    year = "2026",
    journal={},  
}

@article{by8,
    author = "Liu, Jia-Zhou and Wu, Shan-Ping and Wei, Shao-Wen and Liu, Yu-Xiao",
    title = {{Black Hole Entropy Beyond the Wald Term in Nonminimally Coupled Gravity: A Covariant Phase Space Decomposition}},
    eprint = "2605.22429",
    archivePrefix = "arXiv",
    primaryClass = "gr-qc",
    month = "5",
    year = "2026",
    journal={},  
}

@article{by9,
  author = "Liu Wentao and Fang Xiongjun and Jing Jiliang and Wang Jieci",
  title = {{Lorentz violation induces isospectrality breaking in Einstein--bumblebee gravity theory}},
  journal = "Sci. China Phys. Mech. Astron.",
  year = "2024",
  volume = "67",
  number = "8",
  pages = "280413-",
  doi = {10.1007/s11433-024-2405-y},
}

@article{by10,
  author = "Xia Zhong-Wu and Long Sheng and Gong Huajie and Pan Qiyuan and Jing Jiliang",
  title = {{Scalar perturbation around a rotating Kalb--Ramond BTZ black hole}},
  journal = "Sci. China Phys. Mech. Astron.",
  year = "2026",
  volume = "69",
  number = "6",
  pages = "260411-",
  doi = {10.1007/s11433-025-2921-y},
}

@article{by11,
  author       = {Yang, S.J. and Wu, S.P. and Wei, S.W. and others},
  journal      = {Sci. China Phys. Mech. Astron.},
  title        = {{Deciphering black hole phase transitions through photon spheres}},
  year         = {2025},
  volume       = {68},
  pages        = {120412},
  doi          = {10.1007/s11433-025-2787-4},
}

@article{by12,
  author       = {Chen, Z.Q. and Wei, S.W.},
  journal      = {Eur. Phys. J. C},
  title        = {{Thermodynamical topology with multiple defect curves for dyonic AdS black holes}},
  year         = {2024},
  volume       = {84},
  pages        = {1294},
  doi          = {10.1140/epjc/s10052-024-13620-w},
}

@article{by13,
title = {{Topological classes of thermodynamics of the rotating charged AdS black holes in gauged supergravities}},
journal = {Physics Letters B},
volume = {856},
pages = {138919},
year = {2024},
issn = {0370-2693},
doi = {10.1016/j.physletb.2024.138919},
url = {https://www.sciencedirect.com/science/article/pii/S0370269324004775},
author = {Xiao-Dan Zhu and Di Wu and Dan Wen},
}

@article{by14,
  title = {{Novel topological classes in black hole thermodynamics}},
  author = {Wu, Di and Liu, Wentao and Wu, Shuang-Qing and Mann, Robert B.},
  journal = {Phys. Rev. D},
  volume = {111},
  issue = {6},
  pages = {L061501},
  numpages = {8},
  year = {2025},
  month = {Mar},
  publisher = {American Physical Society},
  doi = {10.1103/PhysRevD.111.L061501},
  url = {https://link.aps.org/doi/10.1103/PhysRevD.111.L061501},
}

@article{by15,
doi = {10.1088/1674-1137/ad57b0},
url = {https://doi.org/10.1088/1674-1137/ad57b0},
year = {2024},
month = {sep},
volume = {48},
number = {9},
pages = {095109},
author = {Wang, Han and Du, Yun-Zhi},
title = {{Topology of charged AdS black hole in restricted phase space*}},
journal = {Chinese Physics C},
}

@article{by16,
doi = {10.1088/1674-1137/ad711b},
url = {https://doi.org/10.1088/1674-1137/ad711b},
year = {2024},
month = {nov},
volume = {48},
number = {11},
pages = {115115},
author = {Sadeghi, Jafar and Gashti, Saeed Noori and Alipour, Mohammad Reza and Afshar, Mohammad Ali S.},
title = {{Thermodynamic topology of quantum corrected AdS-Reissner-Nordstrom black holes in Kiselev spacetime}},
journal = {Chinese Physics C},

}

@article{by17,
  title = {{Thermodynamical topology of quantum BTZ black hole}},
  author = {Wu, Shan-Ping and Wei, Shao-Wen},
  journal = {Phys. Rev. D},
  volume = {110},
  issue = {2},
  pages = {024054},
  numpages = {14},
  year = {2024},
  month = {Jul},
  publisher = {American Physical Society},
  doi = {10.1103/PhysRevD.110.024054},
  url = {https://link.aps.org/doi/10.1103/PhysRevD.110.024054},
}

@article{by18,
title = {{Notes on thermodynamics of Schwarzschild--like bumblebee black hole}},
journal = {Physics of the Dark Universe},
volume = {45},
pages = {101520},
year = {2024},
issn = {2212-6864},
doi = {10.1016/j.dark.2024.101520},
url = {https://www.sciencedirect.com/science/article/pii/S221268642400102X},
author = {Yu-Sen An},

}

@article{by19,
  author       = {Hu, P.J. and Ma, L. and L\"{u}, H. and others},
  journal      = {Sci. China Phys. Mech. Astron.},
  title        = {{Improved Reall--Santos method for AdS black holes in general 4-derivative gravities}},
  year         = {2024},
  volume       = {67},
  pages        = {280412},
  doi          = {10.1007/s11433-024-2398-1},
}

@article{by20,
title = {{Generalized free energy landscapes from Iyer--Wald formalism}},
journal = {Physics of the Dark Universe},
volume = {51},
pages = {102210},
year = {2026},
issn = {2212-6864},
doi = {10.1016/j.dark.2025.102210},
url = {https://www.sciencedirect.com/science/article/pii/S2212686425004029},
author = {Shan-Ping Wu and Yu-Xiao Liu and Shao-Wen Wei},
}

@article{by21,
  author       = {Ahmed, F. and Silva, E.O.},
  journal      = {Eur. Phys. J. Plus},
  title        = {{Thermodynamic geometry of charged AdS black holes with a string cloud in Lorentz-violating Einstein--Kalb--Ramond gravity}},
  year         = {2026},
  volume       = {141},
  pages        = {709},
  doi          = {10.1140/epjp/s13360-026-07854-8},
}

@article{by22,
title = {{4D AdS Einstein--Gauss--Bonnet black hole endowed with Lorentzian noncommutativity: P--V criticality, Joule--Thomson expansion, and shadow}},
journal = {Annals of Physics},
volume = {458},
pages = {169451},
year = {2023},
issn = {0003-4916},
doi = {10.1016/j.aop.2023.169451},
url = {https://www.sciencedirect.com/science/article/pii/S0003491623002531},
author = {H. Lekbich and A. {El Boukili} and N. Mansour and M.B. Sedra},
}

@article{by23,
  author       = {Deng, W. and Liu, W. and Xiao, K. and others},
  journal      = {Eur. Phys. J. C},
  title        = {{Quasinormal modes of scalar, electromagnetic, and gravitational perturbations in slowly rotating Kalb--Ramond black holes}},
  year         = {2026},
  volume       = {86},
  pages        = {232},
  doi          = {10.1140/epjc/s10052-026-15470-0},
}

@article{by24,
doi = {10.1088/1475-7516/2025/11/028},
url = {https://doi.org/10.1088/1475-7516/2025/11/028},
year = {2025},
month = {nov},
publisher = {IOP Publishing},
volume = {2025},
number = {11},
pages = {028},
author = {Deng, Weike and Liu, Wentao and Long, Fen and Xiao, Kui and Jing, Jiliang},
title = {{Quasinormal modes of a massive scalar field in slowly rotating Einstein--Bumblebee black holes}},
journal = {Journal of Cosmology and Astroparticle Physics},
}

@article{by25,
  author       = {Liu, X. and Liu, W. and Liu, Z. and others},
  journal      = {J. High Energy Phys.},
  title        = {{Harvesting correlations from BTZ black hole coupled to a Lorentz-violating vector field}},
  year         = {2025},
  volume       = {2025},
  number       = {8},
  pages        = {94},
  doi          = {10.1007/JHEP08(2025)094},
}

@article{by26,
  author       = {Tang, Y. and Liu, W. and Wang, J.},
  journal      = {Eur. Phys. J. C},
  title        = {{Observational signature of Lorentz violation in acceleration radiation}},
  year         = {2025},
  volume       = {85},
  pages        = {1108},
  doi          = {10.1140/epjc/s10052-025-14797-4},
}

@article{by27,
doi = {10.1088/1475-7516/2025/05/017},
url = {https://doi.org/10.1088/1475-7516/2025/05/017},
year = {2025},
month = {may},
publisher = {IOP Publishing},
volume = {2025},
number = {05},
pages = {017},
author = {Liu, Wentao and Wu, Di and Wang, Jieci},
title = {{Shadow of slowly rotating Kalb--Ramond black holes}},
journal = {Journal of Cosmology and Astroparticle Physics},
}

@article{by28,
doi = {10.1088/1475-7516/2024/09/017},
url = {https://doi.org/10.1088/1475-7516/2024/09/017},
year = {2024},
month = {sep},
publisher = {IOP Publishing},
volume = {2024},
number = {09},
pages = {017},
author = {Liu, Wentao and Wu, Di and Wang, Jieci},
title = {{Static neutral black holes in Kalb--Ramond gravity}},
journal = {Journal of Cosmology and Astroparticle Physics},
}

@article{by29,
  author       = {Liu, W. and Fang, X. and Jing, J. and others},
  journal      = {Eur. Phys. J. C},
  title        = {{QNMs of slowly rotating Einstein--Bumblebee black hole}},
  year         = {2023},
  volume       = {83},
  pages        = {83},
  doi          = {10.1140/epjc/s10052-023-11231-5},
}

@article{by30,
  author       = {Liu, J.Z. and Guo, W.D. and Wei, S.W. and others},
  journal      = {Eur. Phys. J. C},
  title        = {{Charged spherically symmetric and slowly rotating charged black hole solutions in bumblebee gravity}},
  year         = {2025},
  volume       = {85},
  pages        = {145},
  doi          = {10.1140/epjc/s10052-025-13859-x},
}
\end{document}